\documentclass[fleqn,usenatbib]{mnras}

\usepackage{newtxtext,newtxmath}

\usepackage[T1]{fontenc}
\usepackage{ae,aecompl}
\usepackage{todonotes}

\usepackage{float}
\usepackage{graphicx}	
\usepackage{amsmath}	
\usepackage{amssymb,color}	
\usepackage{natbib}
\usepackage{siunitx}
\makeatletter
\providecommand\add@text{}
\newcommand\tagaddtext[1]{%
  \gdef\add@text{#1\gdef\add@text{}}}%
\renewcommand\tagform@[1]{%
  \maketag@@@{\llap{\add@text\quad}(\ignorespaces#1\unskip\@@italiccorr)}%
}
\makeatother

\title[The Evolution of Earth's Magnetosphere During the Solar Main Sequence ]{The Evolution of Earth's Magnetosphere During the Solar Main Sequence}

\author[S. Carolan et al.]{
S.~Carolan,$^{1}$\thanks{E-mail: carolast@tcd.ie}
A.~A.~Vidotto$^{1}$, C.~Loesch$^{2}$, P.~Coogan$^{1}$
\\
$^{1}$School of Physics, Trinity College Dublin, College Green, Dublin 2, Ireland\\
$^{2}$Instituto de Educa\c c\~ao Superior de Bras\'ilia - IESB Centro Universit\'ario
Bras\'ilia, DF, Brazil
}

\date{Accepted XXX. Received YYY; in original form ZZZ}

\pubyear{2019}

\begin{document}
\label{firstpage}
\pagerange{\pageref{firstpage}--\pageref{lastpage}}
\maketitle

\begin{abstract}
As a star spins-down during the main sequence, its wind properties are affected. In this work, we investigate how the Earth's magnetosphere has responded to the change in the solar wind. Earth's magnetosphere is simulated using 3D magnetohydrodynamic models that incorporate the evolving local properties of the solar wind. The solar wind, on the other hand, is modelled in 1.5D for a range of rotation rates $\Omega$ from 50 to 0.8 times the present-day solar rotation ($\Omega_\odot$). Our solar wind model uses empirical values for magnetic field strengths, base temperature and density, which are derived from observations of solar-like stars. We find that for rotation rates $\simeq 10 \Omega_\odot$, Earth's magnetosphere was substantially smaller than it is today, exhibiting a strong bow shock. As the sun spins down, the magnetopause standoff distance varies with $\Omega^{-0.27}$ for higher rotation rates (early ages, $\geq 1.4 \Omega_\odot$), and with $\Omega^{-2.04}$ for lower rotation rates (older ages, $< 1.4 \Omega_\odot$). This break is a result of the empirical properties adopted for the solar wind evolution. We also see a linear relationship between magnetopause distance and the thickness of the shock on the subsolar line for the majority of the evolution ($\leq 10 \Omega_\odot$). It is possible that a young fast rotating Sun would have had rotation rates as high as $30$ to $50 \Omega_\odot$. In these speculative scenarios, at $30 \Omega_\odot$, a weak shock would have been formed, but for $50 \Omega_\odot$, we find that no bow shock could be present around Earth's magnetosphere. This implies that with the Sun continuing to spin down, a strong shock would have developed around our planet, and remained for most of the duration of the solar main sequence.
\end{abstract}

\begin{keywords}
planets and satellites: magnetic fields -- planets and satellites: terrestrial planets -- planets and satellites: physical evolution -- MHD -- stars: winds
\end{keywords}



\section{Introduction}\label{sec_intro}

In the solar system, all the giant planets and the Earth have intrinsic magnetic fields. These magnetic fields are largely dipolar, and create cavities, preventing the solar wind from reaching the surface directly \citep[e.g.][]{Bagenal2013}. Recently, there have been  discussions on whether smaller or larger magnetospheres can protect the atmospheres of planets \citep{Strangeway2010, Brain2013, 2013A&G....54a1.25V, Tarduno2014, blackman2018}. Some say that a large magnetosphere would act as a shield from stellar wind particles directly impacting the planet, and the larger the shield, the more protected the atmosphere is against erosion. On the other hand, others say that a larger magnetosphere would have a greater collecting area for stellar wind plasma, which would be channelled towards polar regions. This inflow would generate local heating, which could induce atmospheric escape through polar flows \citep[e.g.][]{Moore2007}. It is also possible that both effects co-exist, but they have different contributions depending on the physical characteristics of the system, such as planetary magnetic field strength and energy of stellar winds. \citet{blackman2018}  suggested that it is the competition between low inflow speeds (from stellar winds) and large collecting areas (from magnetospheres) that define whether planetary magnetospheres can have protective effects in a planet's atmosphere. They found that for the Earth, even though the total amount of solar wind material captured in our magnetosphere is larger than that of a fictitious non-magnetic Earth, our magnetic field still has had a protective effect in protecting our atmosphere. This scenario however could be different in Earth's future, due different solar wind characteristics. 

For close-in exoplanets, it has been suggested that, as well as protecting from the stellar wind, magnetic fields can act to hinder planetary outflows \citep[e.g.,][]{Adams2011, Trammell2011, Khod2015}.  Closed magnetic field lines can trap gas close to the planet, creating ``dead-zones" from which mass is not lost \citep{Khod2015}.  Planetary atmospheres would then escape through open field lines, similar to stellar wind theory \citep{Vidotto2009}. For example, in the numerical study of \citet{Owen2014}, it was found that magnetised exoplanets lose a factor of 4 to 8 less mass than unmagnetised exoplanets, as only a fraction of magnetic field lines remain open and night-side loss is suppressed. The amount of material captured in the dead zones is controlled by the geometry of the magnetic field lines \citep{Khod2015}, giving the structure of the magnetosphere importance in understanding atmospheric escape in close-in exoplanets. The shape and size of the magnetosphere is controlled by the stellar wind, so an understanding of the evolving wind-planet interaction is key before implications can be made to evolving atmospheric escape \citep{Egan2019}.

Short term effects can impact the shape and size of the magnetosphere. Impulsive events, such as coronal mass ejections, may briefly increase the strength of the solar wind impacting the planet, causing a short-term variation in the magnetosphere, its surrounding bow shock, and atmosphere of the planet \citep{Ngwira2014, Nature2016}, which could alter atmospheric escape \citep{Johnstone2019}.

As well as these short-term events, the Sun is known to flip polarity on an 11-year cycle. As was seen by \citet{Das2018}, the direction of the stellar wind's magnetic field can have large implications on the structure of the magnetosphere. The most extreme cases caused by this cyclical polarity flip are the open and closed magnetospheres. These occur when the magnetic field of the  wind and the field lines on the day-side of the planet are aligned and anti-aligned respectively. (These cases are discussed further in Appendix \ref{Zcomp}.) In the case where they are parallel on the day-side (no reconnection), the magnetic field of the wind forces the planet's field lines to remain closed. Conversely, when they are anti-parallel (with reconnection), there is a much greater number of open field lines on the night-side and at the poles of the planets. This could lead to inflow/outflow of material at the polar regions, which has implications on the sustainability of a planets atmosphere as previously mentioned. 

Though these relatively short term cyclical variations are important, the long-term evolution of the solar wind will have a larger effect on Earth's magnetosphere. As we will discuss later on, the magnetosphere is influenced by the conditions of the stellar wind, which depends on the magnetic activity of the star. Since stellar activity declines with both age and rotation \citep{skumanich1972, Ribas2005, Vidotto2014}, the wind of the young sun is believed to have been stronger, which then declined with age (or rotation rate) \citep{dualta2018, dualta2019,pognan2018}. In a numerical study of the interaction between the paleo-Earth ($\sim$3.5~Gyr ago) and the young sun, \citet{Sterenborg2011} concluded that the young Sun's wind would have had easier access to the Earth's surface at that age.

In this paper, we examine the evolution of Earth's magnetosphere over the solar main-sequence lifetime. The novelty of our work is that we couple two sets of simulations: one set of simulations characterises the evolving stellar wind and the other characterises the evolution of Earth's magnetosphere, using as input the results of the former. Our work on the interaction between the evolving solar wind and Earth is relevant for contextualising atmospheric protection in own planet and in other exoplanets, which is likely linked to the evolution of life \citep{2011OLEB...41..503L, 2012EP&S...64..179L, 2013ApJ...770...23Z, blackman2018}. Our paper is presented as following. We first simulate the evolution of the solar wind with time using empirical relations for base temperature, density and magnetic field strength (Section \ref{stellar_wind_section}). We use rotation as a proxy for age, in which case rotation of the young Sun is faster than the current rotation rate. Although our models reproduce observations of mass-loss rates derived for fast rotators, they do not consider wind saturation at very fast rotation. This limitation is further discussed in Section \ref{stellar_wind_section}. We then simulate the interaction between the solar wind and Earth's magnetosphere at different ages (rotations) using 3D numerical simulations (Section \ref{SWMF_section}). We examine the variations of the day-side of Earth's magnetosphere and bow shock with rotation (Section \ref{section_EMIT}). At a very early age, it is still unknown whether the sun was a fast, moderate or slow rotator. We explore the extreme environment around the young Earth in Section \ref{sec_high_omega} and present our conclusions in Section \ref{sec_conclusions}.

\section{Stellar wind modelling}\label{stellar_wind_section}
 For our stellar wind modelling, we use a 1.5D Weber-Davis model \citep{weberdavis1967}.  For that, we use the Versatile Advection Code \citep[VAC, ][]{Toth1996}, based on the version of the code from \citet{johnstone2015}. Our wind model is polytropic, such that the pressure and density are related by $p^{\rm sw} \propto (\rho^{\rm sw})^\alpha$. This  relation is enforced in the model and there is no need for an energy equation to be solved \citep{1999A&A...343..251K}. Here, we adopt a constant polytropic index of $\alpha = 1.05$ in our simulations, which implies that the stellar wind temperature profile, for each model, is nearly isothermal.. Additionally, we consider the stars to be magnetised and rotating. The rotation rate $\Omega$ is varied from $0.8\Omega_\odot$ to $50\Omega_\odot$, to mimic the solar wind evolution through the main-sequence phase. The wind temperature, density and magnetic field depend on rotation in our models. We describe next how we chose these wind parameters.

\subsection{Choice of stellar wind parameters}
Polytropic wind models have two important free parameters, namely the temperature and density at the base of the wind. The values of these parameters are typically assumed to be coronal values, which generate hotter and rarefied winds, similar to what is adopted for the present-day Sun \citep[e.g.][]{1971SoPh...18..258P}. However, these parameters are not easy to measure in stars other than the Sun.
 To derive the wind temperatures of low-mass stars,  theoretical works have either assumed a relationship between temperature and X-ray fluxes \citep[e.g.,][]{Holzwarth2007,  johnstone2015, Reville2016, dualta2018} or being proportional to the square-root of the escape velocity \citep[e.g.,][]{Matt2012}.  
 These two families of models have been discussed in depth by \citet{johnstone2015}. They are, by definition, equivalent for the present-day Sun, but for other stars, they can lead to much different wind models \citep[see also][]{2018haex.bookE..26V}.
In the latter approach, for example, the escape velocity does not vary significantly in the main sequence, during which the stars spin down and become less active. This implies that wind temperatures in these models would be approximately constant throughout the main-sequence evolution. In our models, we use the former approach -- given that X-ray emission is seen to vary by several orders of magnitude for stars at different rotation rates \citep[e.g.,][]{Pizzolato2003}, we naively would expect that a high-temperature corona would lead to a high-temperature wind and, hence, we adopt in our models a correlation between the two. More specifically, we follow the approach by \citet{dualta2018}, who modeled the wind base temperature ($T_0$) by scaling the average coronal temperatures of Sun-like stars to current solar wind values following the X-ray flux--temperature relations of \citet{JohnstoneGudel2015}. Our base temperature is dependent on the stellar rotation rate $\Omega$ according to:
\begin{equation}
  T_0~[K]=\begin{cases}
    1.5\times10^6\big(\frac{\Omega}{\Omega_\odot}\big)^{1.2}& {\rm for} ~~~~~ \Omega < 1.4 \Omega_\odot, \\\\
    1.98\times10^6\big(\frac{\Omega}{\Omega_\odot}\big)^{0.37} & {\rm for} ~~~~~ \Omega \geq 1.4 \Omega_\odot.
  \end{cases}
  \label{BaseTeqn}
\end{equation}
 \citet{dualta2018} fitted a broken power law to the X-ray data in light of other works suggesting a break in other activity quantities  -- rotation rates, lithium abundances and x-ray luminosity \citep{Booth2017, Beck2017, vansaders2016}.

For the base number density ($n_0$) we use a rotation-dependent density relation, derived by \citet{IvanovaTaam2003} from X-ray  observations, and employed in other wind studies of solar-like stars \citep{Holzwarth2007, Reville2016, dualta2018}:
\begin{equation}
    n_0~[g/cm^3] = 10^8 \Big(\frac{\Omega}{\Omega_\odot}\Big)^{0.6}.  
\end{equation}

 In a Weber-Davis model, the magnetic field lines are assumed to be approximately radial at the wind base. In the initial condition of our simulations, we  assume that the field is purely radial and decays with distance-squared. As the simulation evolves, in addition to the  radial magnetic field component $B_r^{\rm sw}$, an azimuthal component $B_{\phi}^{\rm sw}$ is created due to stellar rotation.  The spiral angle that characterises the tightness of the  Parker spiral is $\Psi = \arctan(B_{\phi}^{\rm sw}/{B^{\rm sw}_{r}})$.  The magnetic field strength at the wind base $B_{r,0}^{\rm sw}$ is derived from the empirical relation from \citet{Vidotto2014}, based in observationally-derived magnetic maps, 
\begin{equation}
    B_{r,0}^{\rm sw}~[G] = 1.29 \Big(\frac{\Omega}{\Omega_\odot}\Big)^{1.32}.
\end{equation}
Note that the values from \citet{Vidotto2014} represent an average field strength of the large-scale magnetic field over the stellar surface and, here, is used as the radial component of the field strength.

\begin{figure*}
    \includegraphics[width=\textwidth]{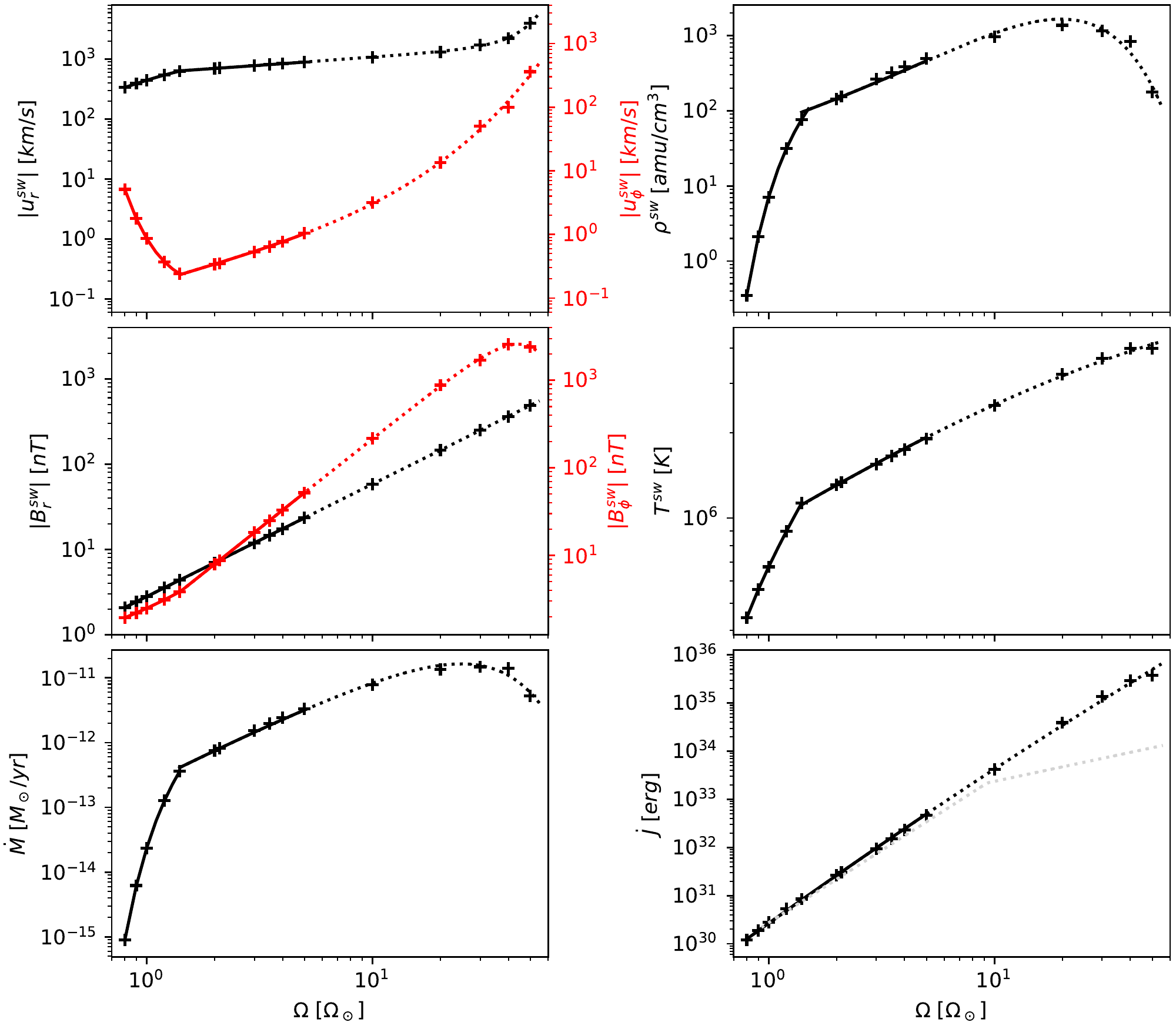}
    \caption{Stellar wind local velocity, density, magnetic field and temperature at 1au, and mass and angular momentum loss rate profiles for stellar rotation rates from 0.8 $\Omega_\odot$ to 50 $\Omega_\odot$. Fits are shown in black and red (when an azimuthal component is shown in the same panel). The fitting parameters are in Tables
    \ref{table_fits1} and \ref{table_fits2}. The crosses mark the results of a particular model and are also listed in Table \ref{wind-table}.  To represent the uncertainty of whether the Sun would have rotated faster than $\sim 5 \Omega_\odot$, part of the fits are shown as dotted lines. The grey dotted line in the last panel shows the break in angular-momentum loss rates one would expect if our models adopted saturation of the wind (see text for a discussion).}
    \label{Wind_4Panel}
\end{figure*}

Our wind simulations extend to 1~au (equivalent to $215R_\odot$). For each one of them, we compute the density, temperature, the radial ($r$) component of velocity $u_r^{\rm sw}$ and magnetic field, as well as the azimuthal ($\phi$) component of velocity $u_\phi^{\rm sw}$ and magnetic field with distance. The values of these quantities at 1 au, for each assumed rotation rate,  are shown in Figure \ref{Wind_4Panel} and listed in Table \ref{wind-table}.

 \begin{table*}
 \caption{The absolute magnitude of wind parameters for different stellar rotation rates $\Omega$. The columns are: estimated age based on the rotation tracks from \citet{Gallet2013}, the mass- ($\dot{M}$)  and angular momentum-loss rates ($\dot{J}$),  the local radial and azimuthal components of the velocity ($u_{r}^{\rm sw}$, $u_{\phi}^{\rm sw}$), magnetic field ($B_{r}^{\rm sw}$, $B_{\phi}^{\rm sw}$), temperature ($T^{\rm sw}$), proton number density ($n_p^{\rm sw}$). $\mathcal{M}$ is the local magnetosonic mach number and $\Psi$ is the spiral angle of the wind. All the local quantities were computed  at 1 au and are used as input in the 3D simulations of Earth's magnetosphere. }
 \label{wind-table}
 \begin{tabular}{cccccccccccccc}
 \hline
 $\Omega$ & Age & $\dot{M}$ & $\dot{J}$  &$u_r^{\rm sw}$ & $u_\phi^{\rm sw}$ & $B_r^{\rm sw}$ & $B_\phi^{\rm sw}$& $T^{\rm sw}$& $n_p^{\rm sw}$ & $\mathcal{M}$ & $\Psi$\\

$[\Omega_\odot]$ & [Myr] & [$M_\odot$/yr] & [erg] &[km/s]& [km/s]&  [G]& [G] & $[\times10^5$ K] & [cm$^{-3}$] &  & [$^\circ$] \\
\hline
	0.8 & 7700 & $8.9\times10^{-16}$ & $1.2\times10^{30}$ & 337 & 5.1 & $2.1\times10^{-5}$ & $1.9\times10^{-5}$ & 4.4 & 0.35 & 2.5 & 47\\
	0.9 & 6500 & $6.2\times10^{-15}$ & $1.9\times10^{30}$ & 393 & 1.8 & $2.4\times10^{-5}$ & $2.2\times10^{-5}$ & 5.6 & 2.1 & 3.6 &  48\\
	1.0 & 5000 & $2.3\times10^{-14}$ & $2.8\times10^{30}$ & 446 & 0.86 & $2.8\times10^{-5}$ & $2.5\times10^{-5}$ & 6.7 & 7.0 & 4.0 & 48\\
	1.2 & 3100 & $1.3\times10^{-13}$ & $5.3\times10^{30}$ & 545 & 0.37 & $3.5\times10^{-5}$ & $3.1\times10^{-5}$ & 9.0 & 32 & 4.3 &49 \\
	1.4 & 2100 & $3.6\times10^{-13}$ & $8.5\times10^{30}$ & 634 & 0.24 & $4.4\times10^{-5}$ & $3.8\times10^{-5}$ & 11 & 76 & 4.5 & 49\\
	2.0 & 1100 & $7.5\times10^{-13}$ & $2.6\times10^{31}$ & 702 & 0.34 & $7.0\times10^{-5}$ & $7.9\times10^{-5}$ & 13 & 143 & 4.6 & 41\\
	5.0 & 40 -- 440 & $3.3\times10^{-12}$ & $4.6\times10^{32}$ & 900 & 1.0 & $2.3\times10^{-4}$ & $5.2\times10^{-4}$ & 19 & 497 & 4.7 & 24\\
	10 & $\leq$280 & $7.8\times10^{-12}$ & $4.2\times10^{33}$ & 1078 & 3.1 & $5.8\times10^{-4}$ & 0.0022 & 25 & 968 & 4.1 & 15\\
	30 & $\leq$140 & $1.5\times10^{-11}$ & $1.4\times10^{35}$ & 1723 & 50.5 & 0.0025 & 0.017 & 37 & 1154 & 1.5 & 8 \\
	50 & $\leq$100 & $5.3\times10^{-12}$ & $3.8\times10^{35}$ & 3994 & 360 & 0.0049 & 0.024 & 40 & 177 & 0.99 & 11\\
      \hline
\end{tabular}
\end{table*}

These wind parameters at 1 au are the inputs of our magnetosphere simulations (cf.~Section \ref{SWMF_section}). The magnetosonic velocity at the interaction, together with the stellar wind velocity,  determine the  magnetosonic mach number 
\begin{equation}
    \mathcal{M} = \frac{u_r^{\rm sw}}{\sqrt{v_A^2 + c_s^2}},
\end{equation}
where $v_A$ is the Alfven velocity and $c_s$ is the sound speed.   $\mathcal{M}$ determines the strength of the shock. Given that a shock is only present around the Earth if $\mathcal{M}>1$, we note a surprising result that, at rotation rate of $50\Omega_\odot$, our models do not predict the presence of a shock around the Earth. As we will see next, it is  uncertain whether the Sun rotated that fast -- if that indeed occurred, it happened only for a short amount of time, relative to the Sun's lifetime, and at an age $\lesssim 100$~Myr. We  discuss age determination and rotation-age relation next.

\subsection{The rotation rate of the young Sun}
\begin{figure}
    \centering
    \includegraphics[width=\columnwidth]{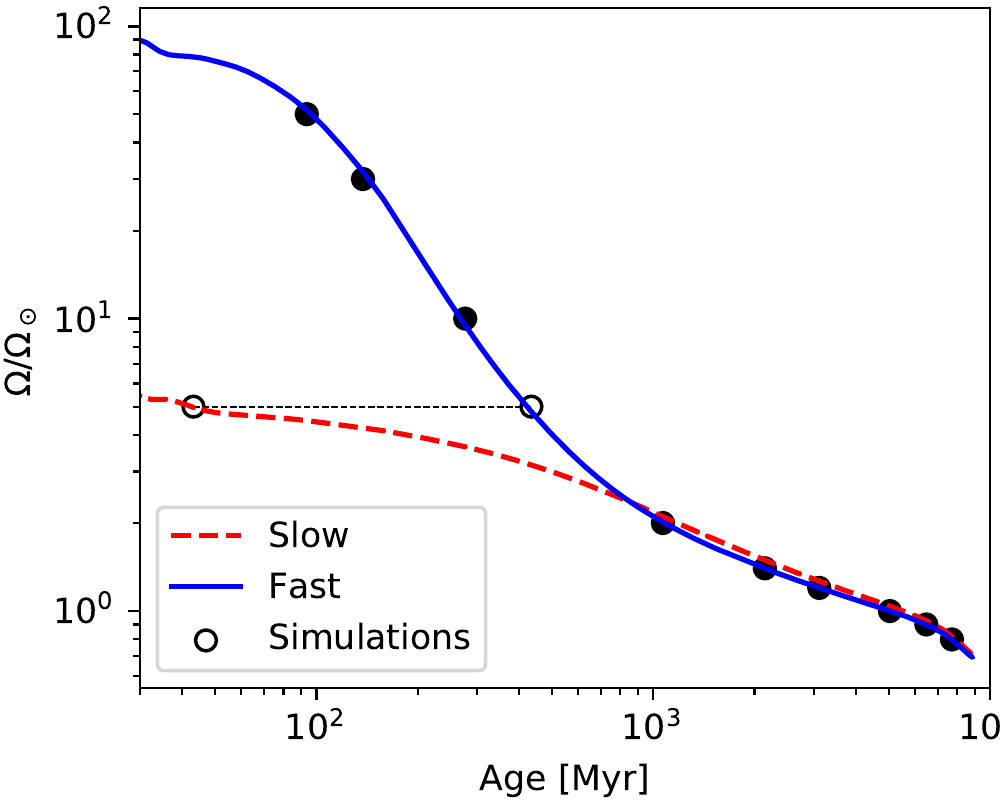}
    \caption{The evolution of stellar rotation rate ($\Omega$) from \citet{Gallet2013}, for a $1-M_\odot$ star. The blue solid line tracks the evolution of a fast rotating solar like star, while the dashed red line tracks the slow rotator. The black points mark the values of $\Omega$ adopted in our simulations. The fast and slow models give a well constrained age for the $\lesssim 2.0~\Omega_\odot$ models. For higher rotation rates, ages are more uncertain. For example, the Sun could have rotated at   $5~\Omega_\odot$ at any age between 40 and 440~Myr, depending whether it was a slow to fast rotator (connected by a dashed black line). Note that the Sun would only have rotated  faster than $\sim 5~\Omega_\odot$, if it were not born as a slow rotator.}
    \label{fig_gallet}
\end{figure}

To mimic the ageing of the Sun, we use rotation as a proxy for age, and compute stellar wind models at different rotation rates from $0.8 \Omega_\odot$ to $50 \Omega_\odot$. However, here, we  are very careful at assigning an age to our models as the age-rotation relationship is only well constrained for stars older than $\sim 800$~Myr. Figure \ref{fig_gallet} shows the rotation evolution tracks for a 1-$M_\odot$ star during the main-sequence phase, extracted from the work of \citet{Gallet2013}. These tracks are the upper/lower envelopes of the observed rotation distributions from open clusters and they indicate the evolution of a slowly-rotating (red-dashed line) and a fast-rotating (solid blue) star. The convergence of the slow  and fast rotator tracks happens at  around $\sim 800$~Myr, after which, we can assign an age to a star from observed rotation rates -- this forms the basis of the gyrochronology method \citep{2003ApJ...586..464B}.

The symbols in Figure \ref{fig_gallet} represent the selected rotation rates to which we perform our simulations. As can be seen, for  $\Omega \lesssim 2~\Omega_\odot$, there is a unique function between age and rotation and thus we can assign an age to our solar wind models. However, for $\Omega \gtrsim 2~\Omega_\odot$, there are ambiguities in age determination. For example, for the model with rotation rate of $5~\Omega_\odot$ (open symbols, connected by a dotted line), possible ages  range from 40 to 440~Myr, depending whether the Sun used to be a slow or a fast rotator, respectively. Note that the Sun would only have rotated  faster than $\sim 5~\Omega_\odot$, if it were not born as a slow rotator. Because of this, it is only possible to assign an upper limit on the age for the models with a given rotation rate  $ \gtrsim 2~\Omega_\odot$ --  for that, we assume that the Sun was in the fast rotator track. These age estimates and upper limits are listed in Table \ref{wind-table} for each of our simulated rotation rates.

Given that we are unsure whether the Sun indeed rotated faster than $\sim 5~\Omega_\odot$, the computed quantities associated to these rotation rates are represented as dotted lines in Figure \ref{Wind_4Panel}. The very fast rotating young Sun ($30$ and $50~\Omega_\odot$) will be discussed in Section \ref{sec_high_omega}.

\subsection{Global properties of the young solar wind}
 In addition to the local conditions (at 1au) of the solar wind in time, we also present in the bottom panels of Figure \ref{Wind_4Panel} and Table \ref{wind-table} two global quantities of stellar winds -- the mass-loss rate
\begin{equation}
    \dot{M} = 4\pi ({\rm 1~au})^2 (m_p n_p^{\rm sw}) u_r^{\rm sw} 
\end{equation}
and the angular-momentum loss rate
\begin{equation}
    \dot{J} = \frac{2}{3} \dot{M} r_A^2 \Omega \, .
\end{equation}
Here, $m_p$ is the mass of the proton and $r_A$ is the distance where the wind radial velocity reaches the Alfven velocity, and is known as the Alfven point (or Alfven surface in a multi-dimension wind model). We discuss the evolution of $\dot{M}$ and $\dot{J}$ next.

Mass-loss rates of solar type stars are challenging to derive from observations. The most successful technique to date has been the astrospheric Lyman-$\alpha$ absorption \citep{Wood2014, Wood2005}, which has derived mass-loss rates for about a dozen stars. This technique has shown that mass-loss rates increase with X-ray fluxes ($F_x$), which also increase with $\Omega$.  For ages $\gtrsim 1000$ Myr, the mass loss rate in our models are very similar to those predicted by \citet{Wood2014} -- this covers the majority of our models ($\leq10~\Omega_\odot$). For example, our models predict that the Sun had a mass-loss rate of $7.5\times10^{-13}$~$M_\odot$/yr, or about  40 times the present-day solar mass-loss rate, at a rotation rate of $2 \Omega_\odot$. At a similar rotation rate, \citet{2002ApJ...574..412W} found that the star $\epsilon$~Eri shows a mass loss rate of $\sim 6 \times 10^{-13} M_\odot/$yr, similar to the value we find in our models.

For $F_x \gtrsim 10^6~$erg~cm$^{-2}$~s$^{-1}$, i.e., above approximately $15~\Omega_\odot$, \citet{Wood2014} predicted a break in the $\dot{M}$--$F_x$ relation, arguing that mass-loss rates of a very young Sun (high $F_x$) could actually have been very low ( $\sim 10^{-14} M_\odot/$yr). This break, however, has been difficult to explain with other observations \citep[e.g.,][]{Vidotto2016}, and theoretically \citep[e.g.][]{Holzwarth2007, 2014A&A...570A..99S}. It has been suggested that the low mass-loss rates of the `outlier' stars in Wood's sample ($\pi^1$~UMa, $\xi$~Boo A)  do not follow the $\dot{M}$--$F_x$ relation due to scattering, and that the relation between mass-loss rates and activity could extend to higher X-ray fluxes without a break \citep{2019MNRAS.482.2853J}, albeit with some scatter. 

\citet{2019MNRAS.482.2853J} estimated that  AB Dor, a widely used proxy for the young Sun, has mass-loss rate  $7\times 10^{-12}~M_\odot$/yr. With a rotation period of $\sim 0.5$~day, or $54 \Omega_\odot$, their derived mass-loss rate is surprisingly similar to our $50 \Omega_\odot$ model ($5.3\times 10^{-12}~M_\odot$/yr). Another clue that fast-rotating stars might actually have stronger stellar winds is that when we go to even younger stars (not studied in this work), mass-loss rates are observed to be $\sim 10^{-10}$ to $10^{-12}~M_\odot$/yr in the weak T Tauri phase, after disc clearing \citep{VidottoDonati2017}. Naively, one would expect that the wind of a young sun would have mass-loss rates that are intermediate between those of weak T Tauri stars and the current solar one. Altogether, these facts give support to the relatively high mass-loss rates, reaching $10^{-11}M_\odot$/yr, we obtain for very fast rotating Suns.

Our model, however, does not include  saturation on the mass-loss rate nor on magnetic field at high rotation rates\footnote{ Although we do not impose a saturation in mass-loss rate, our models show a `levelling off' of mass-loss rate for $\Omega \gtrsim 20\Omega_\odot$ (cf.~the bottom left panel of  Figure \ref{Wind_4Panel}). This inflection is more clearly seen in the local densities (top right panel of  Figure \ref{Wind_4Panel}). Although the base densities are larger for larger rotation rates, the decrease in wind density with distance  is  steeper for $ \gtrsim 20\Omega_\odot$. In the limit where the wind reaches terminal velocity, the density should fall with $r^{-2}$. The steeper decrease is thus an indication that the wind is still being accelerated for $\Omega \gtrsim 20 \Omega_\odot$ at 1 au. In fact, we note that in these cases, the local magnetocentrifugal force (at 1au) is comparable or greater than the thermal pressure gradient in the wind radial momentum equation. Beyond $\gtrsim 20 \Omega_\odot$, the thermal pressure gradient ceases to be the dominant force in our models.}. Saturation is required  to explain the spin down of the very fast rotators \citep[e.g.][]{2015ApJ...799L..23M, 2015A&A...577A..28J}. When considering saturation, the  wind angular momentum-loss rate presents a break with $\Omega$, which is illustrated by the grey dotted line in the bottom right panel of Figure \ref{Wind_4Panel}. This curve is from See et al (submitted), which is based on the torque formalism of  \citet[][we divided their curve by 2.9 to match our solar value]{2015ApJ...799L..23M}. We note that for $\Omega \lesssim 10 \Omega_\odot$, their trend is similar to ours, roughly obeying a cubic dependence with $\Omega$. However, for $\Omega \gtrsim 10 \Omega_\odot$, saturation requires an approximately linear dependence between $\dot{J}$ and $\Omega$, which is not seen our model.

\section{Magnetosphere Modeling} \label{SWMF_section}
There have been numerical studies investigating present-day and past interactions between the solar wind and solar system planets, such as with Mars \citep{2009AsBio...9...55T, Ma2013, 2015GeoRL..42.9113M,Sakai2018}, Venus \citep{2007SSRv..129..207K,2009JGRA..114.9208T} and Earth \citep{Ridley2001, vogt2004, Sterenborg2011}. However, to the best of our knowledge, the magnetospheric evolution of the Earth over the solar main sequence lifetime has not yet been examined. We investigate this evolution through 3D magnetohydrodynamic (MHD) modelling of Earth's magnetosphere. We use the Space Weather Modelling Framework's (SWMF) Global Magnetosphere module \citep{Toth-swmf}. SWMF has been used to study planets in the solar system \citep[e.g.,][]{Sterenborg2011,Ma2013, Jia2015, Jia2016}. 

\begin{figure}
	\includegraphics[width=\columnwidth]{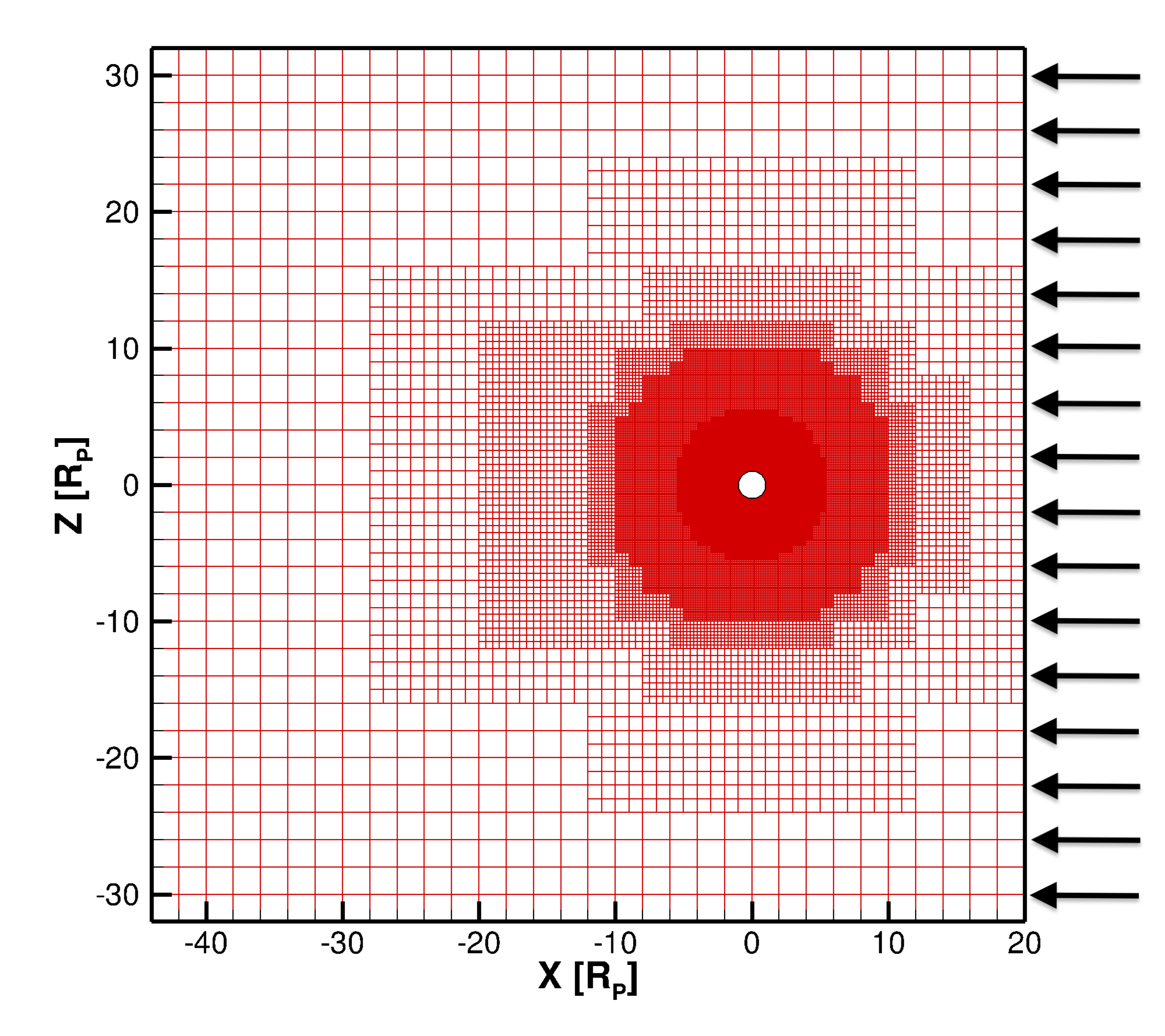}
    \caption{The refinement grid used in our models. A maximum resolution of $1/32$~$R_p$ is used within a radius of 5~$R_p$, which slowly decreases outwards by a factor of 2 at each step as seen above. The arrows visualize the injection of the stellar wind into this domain.}
    \label{fig_grid}
\end{figure}

To simulate Earth's magnetosphere and its surrounding bow shock in a given stellar wind condition, our magnetosphere simulations take the stellar wind parameters as  inputs. The wind is injected into our computation domain at a distance of 20 planetary radii ($R_p$) on the dayside of the planet, as shown in Figure \ref{fig_grid}. For the planetary parameters, we use current day values for magnetic dipole strength ($B_0=-0.3~$G); radius  ($R_{p}=6.3\times10^8~$cm); and mass ($5.976\times10^{27}~$g) in our models. We assume that the Earth's geodynamo has not changed during this evolution, although some works suggest that Earth's dipolar field strength might have been 50\% smaller $\sim$ 3.5 Gyr ago \citep{Tarduno2010}. We discuss the effects our hypothesis of constant dipolar field strength has on our simulations in Section \ref{section_EMIT}.

Our simulation is Cartesian and solves for eight parameters: mass density ($\rho$), velocity ($u_x, u_y, u_z$), magnetic field ($B_x, B_y, B_z$), and  thermal pressure ($P_T$). These are found through iteratively solving a set of ideal MHD equations: the mass conservation, momentum conservation, magnetic induction and energy conservation equations, respectively:
\begin{equation}
    \frac{\partial \rho}{\partial t} + \nabla \cdot (\rho \Vec{u}) = 0,
\end{equation}
\begin{equation}
    \frac{\partial(\rho\Vec{u})}{\partial t} + \nabla \cdot \Bigg[\rho \Vec{u} \Vec{u} + (P_T + \frac{B^2}{8\pi})I - \frac{\Vec{B}\Vec{B}}{4\pi}\Bigg] = \rho \Vec{g}, 
\end{equation}
\begin{equation}
    \frac{\partial\Vec{B}}{\partial t} + \nabla \cdot (\Vec{u}\Vec{B} - \Vec{B}\Vec{u}) = 0,
\end{equation}
\begin{equation}
    \frac{\partial \epsilon}{\partial t} + \nabla \cdot \Bigg[\Vec{u}\Bigg(\epsilon + P_T + \frac{B^2}{8\pi}\Bigg)-\frac{(\Vec{u}\cdot\Vec{B})\Vec{B}}{4\pi}\Bigg] = \rho \Vec{g} \cdot \Vec{u},
\end{equation}
where I denotes the identity matrix and $\Vec{g}$ the acceleration due to gravity. The total energy density $\epsilon$ is 
\begin{equation}
    \epsilon = \frac{\rho u^2}{2} + \frac{P_T}{\gamma -1} + \frac{B^2}{8\pi}, 
\end{equation}
with {$\gamma=5/3$}.

We place the centre of the planet at the origin, within a rectangular box. Since we are considering the dayside of the planet in this paper, we choose a cubic grid of length 32~$R_p$ ($x=[-44, 20]$ $R_p$ ; $y=z=[-32, 32]$ $R_p$) as seen in Figure \ref{fig_grid}. The X axis points towards the star and the Z axis is oriented perpendicular to the ecliptic plane. The Y axis constructs the right-handed system. Our simulations have a maximum resolution of $1/32$~$R_p$ within a radius of $5$~$R_p$, which gradually decreases to a minimum resolution of $2$~$R_p$ at the edges of the grid. We can achieve this high resolution by sacrificing grid space on the night side of the planet. We tested the effects of numerical resolution, by changing the highest resolution from $1/32~R_P$ to $1/64~R_P$ and $1/128~R_P$ at the inner regions of our simulations. We found no significant change in the position of pressure balances (used to identify the magnetopause, further discussed in Section \ref{section_EMIT}) and so in this work, we present the results for the case of $1/32~R_p$. With this resolution, our simulations contain 22.6 million cells.

The inner boundary is set at $1R_p$ in our simulations. We chose values for base density ($10~amu/cm^3$), temperature ($25000~K$) and thermal pressure ($3.45\times10^{-11}~dyn/cm^2$) that are appropriate for the current ionosphere and keep them the same for all our simulations. Note that the ionospheric structure is not computed in our models.  Since we are interested in the interaction region of the magnetosphere with the stellar wind, which happens significantly above the planet, the values of pressure and density have no effects on the dynamics of the interaction (for example, for different values of density, we see no change in the position of thermal-magnetic and thermal-ram pressure balances). The boundary assumes that the density at $1R_p$ is fixed, and the magnetic field and thermal pressure have outflow conditions. The velocity vector is reflected upon reaching this boundary in the frame corotating with the planet. The outer boundary assumes an outflow of all parameters. Earth's rotation is kept at 1 day for all the simulations. 

\begin{figure}
    \centering
	\includegraphics[width=0.75\columnwidth]{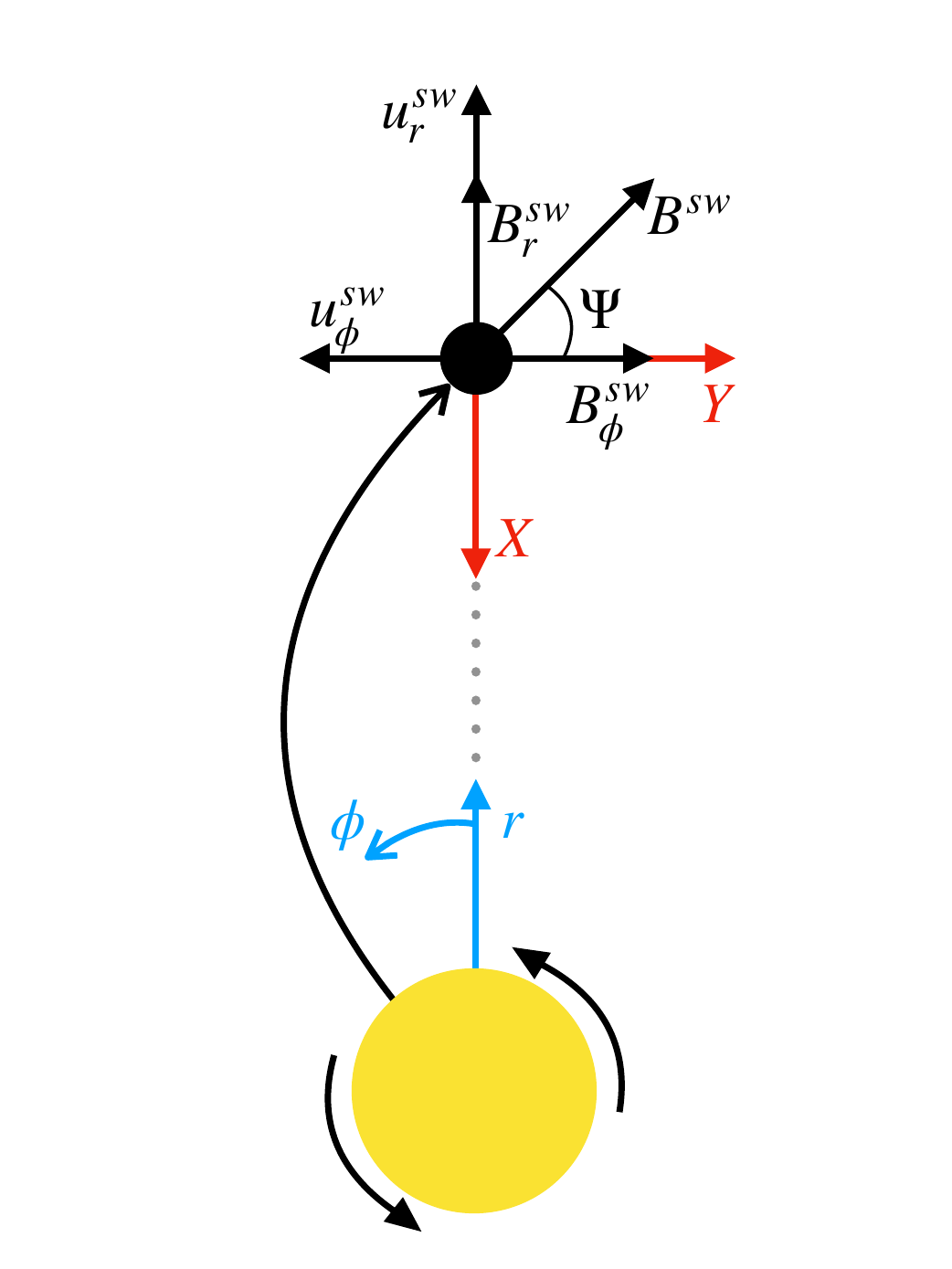}
    \caption{Illustration of how the stellar winds are oriented in our grid. The black circle represents the Earth, while the yellow represents the Sun. The magnetic and velocity vectors are drawn in the inertial frame of the star. The blue vectors show the stellar coordinate system, while the red show the orientation of the planetary coordinate system. Z constructs the right handed system, with the origin placed at the centre of the planet.}
    \label{fig_windorientation}
\end{figure}

In our simulations we align the magnetic axis of the Earth with the rotational axis, which is $23.5^\circ$ inclined with respect to the orbital plane. All the simulations are thus at summer solstice. Presently, the Earth's magnetic axis is misaligned by $\sim 11^\circ$ to the rotational axis. However, as the magnetic tilt can vary on a relatively short timescale compared to the evolutionary times considered here, and given that we do not know precisely how this variation would have happened in the distant past, we chose to neglect the misalignment between magnetic and rotation axes in this work. The radial velocity $u_r^{\rm sw}$ of the solar wind is injected along the negative X direction in this coordinate system. $u^{\rm sw}_\phi$ is aligned with the negative Y direction such that the simulated 1.5D winds act in the X-Y plane. Note that we assume $B_r^{\rm sw}>0$. Hence $B_\phi^{\rm sw}<0$ in the stellar reference frame due to the trailing nature of the Parker spiral. This is illustrated in Figure \ref{fig_windorientation}.

\begin{figure}
	\includegraphics[width=\columnwidth]{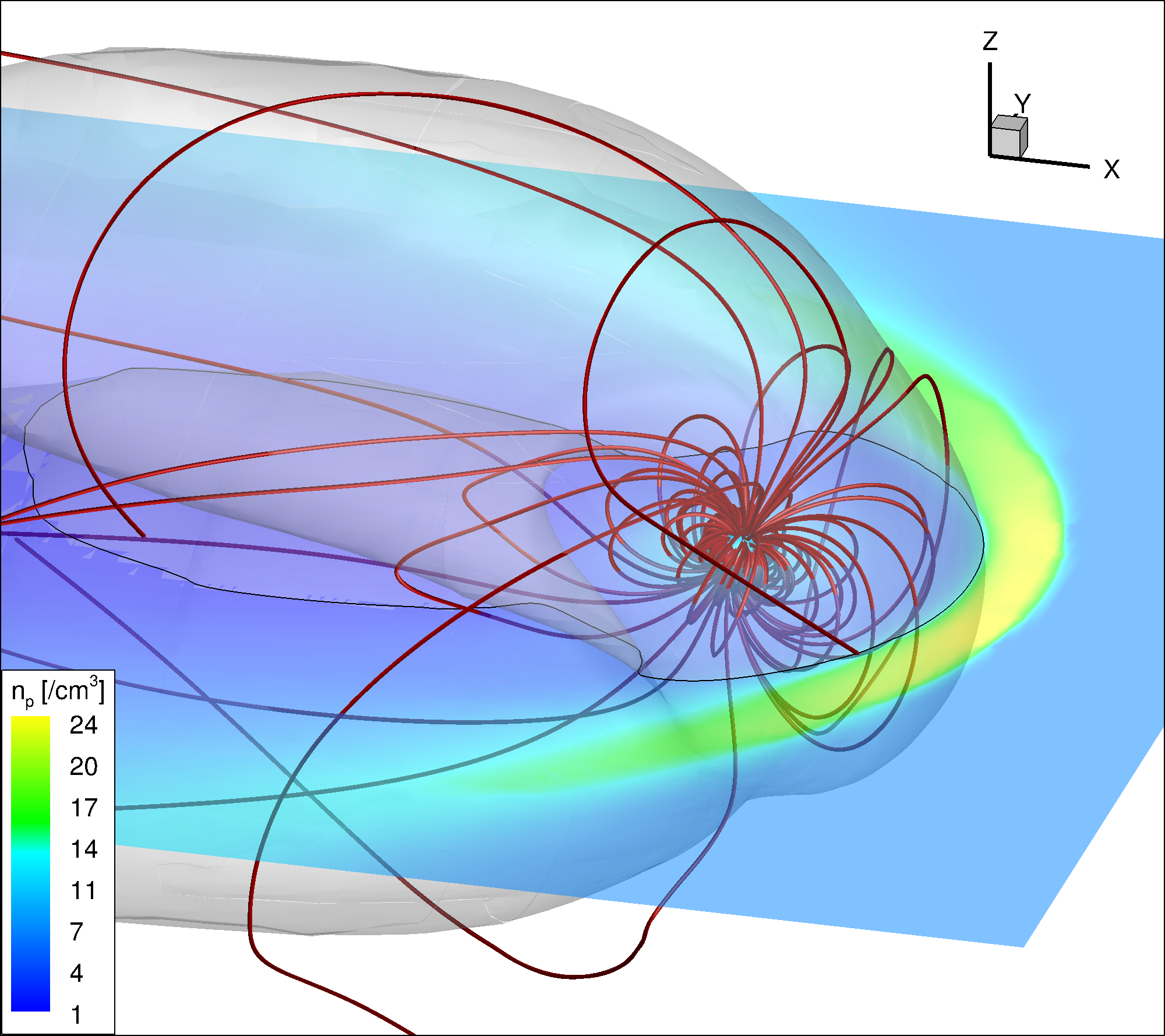}
    \caption{A 3D image of the planet considering the star is rotating at $1.0$~$\Omega_\odot$. Red lines represent the magnetic field lines connected to the planet. The slice shows the density around the planet in the X-Y plane. The surface marks a constant pressure value, which is where the thermal pressure equals the magnetic pressure. We use this surface to mark the magnetopause boundary on the dayside of the planet in our models.}
    \label{fig_3d}
\end{figure}

As $u_r^{\rm sw} \gg u_\phi^{\rm sw}$ the wind is mainly radial at 1 au. This orientation yields a solar wind which enters our domain through the day-side of the box seen in Figure \ref{fig_grid} where it interacts with the planet's magnetic field. Once the wind reaches the magnetic field from the planet, it is shocked which is visualized in Figure \ref{fig_3d} by the density slice in the X-Y plane. The magnetic field from the planet deflects most incoming material.

\section{Earth in Time}\label{section_EMIT}

As stellar rotation is a proxy for age \citep{skumanich1972}, here we choose to sample 10 stages in the solar wind's evolution spanning a range of stellar rotation rates from $0.8 \Omega_\odot$ to $50 \Omega_\odot$ as seen in Table \ref{wind-table}. The magnetosphere and bow shock formed from the interaction between these winds and the Earth's magnetic field can be seen in Figure \ref{fig_2d_1}.

\begin{figure*}
    \includegraphics[width=\columnwidth]{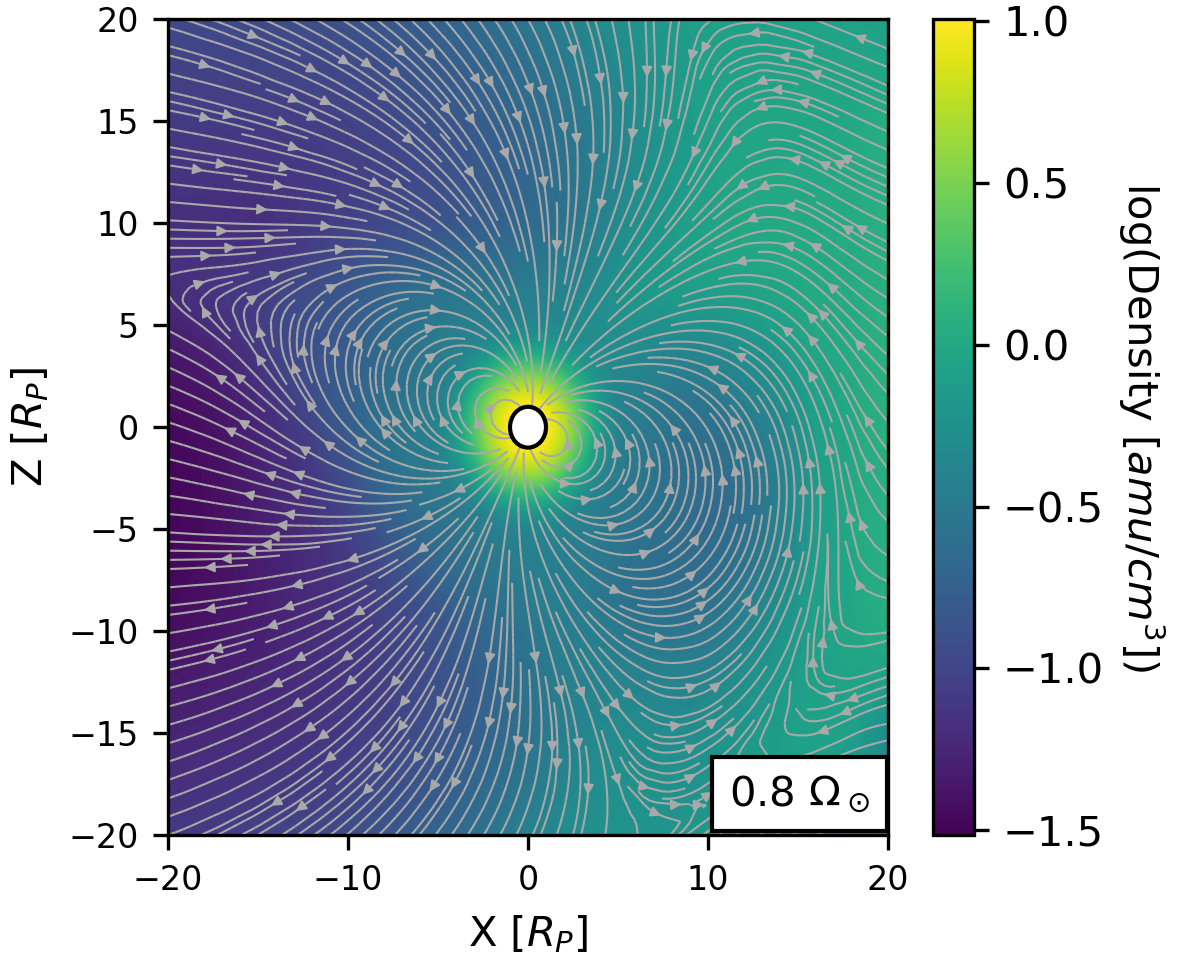}
    \includegraphics[width=\columnwidth]{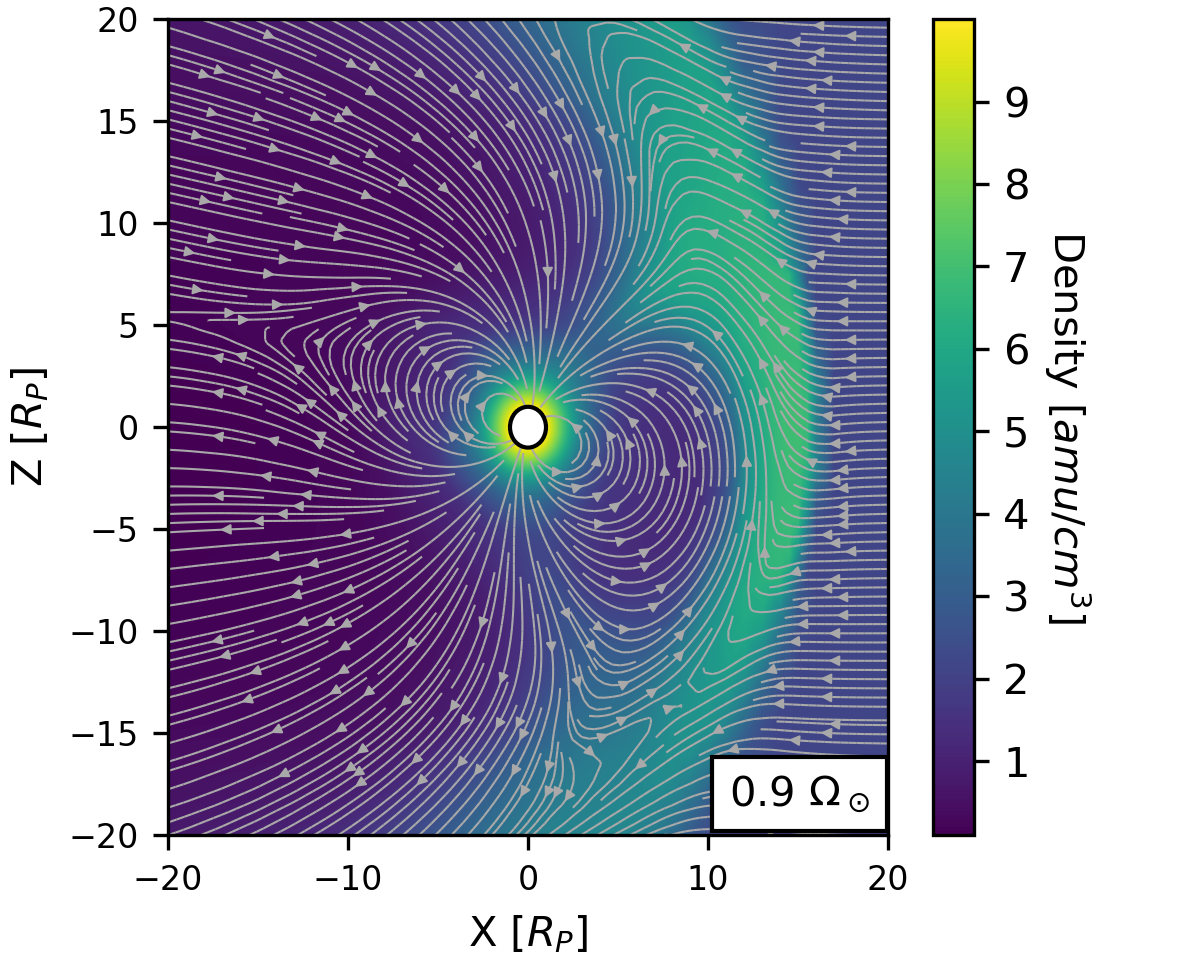}\\
    \includegraphics[width=\columnwidth]{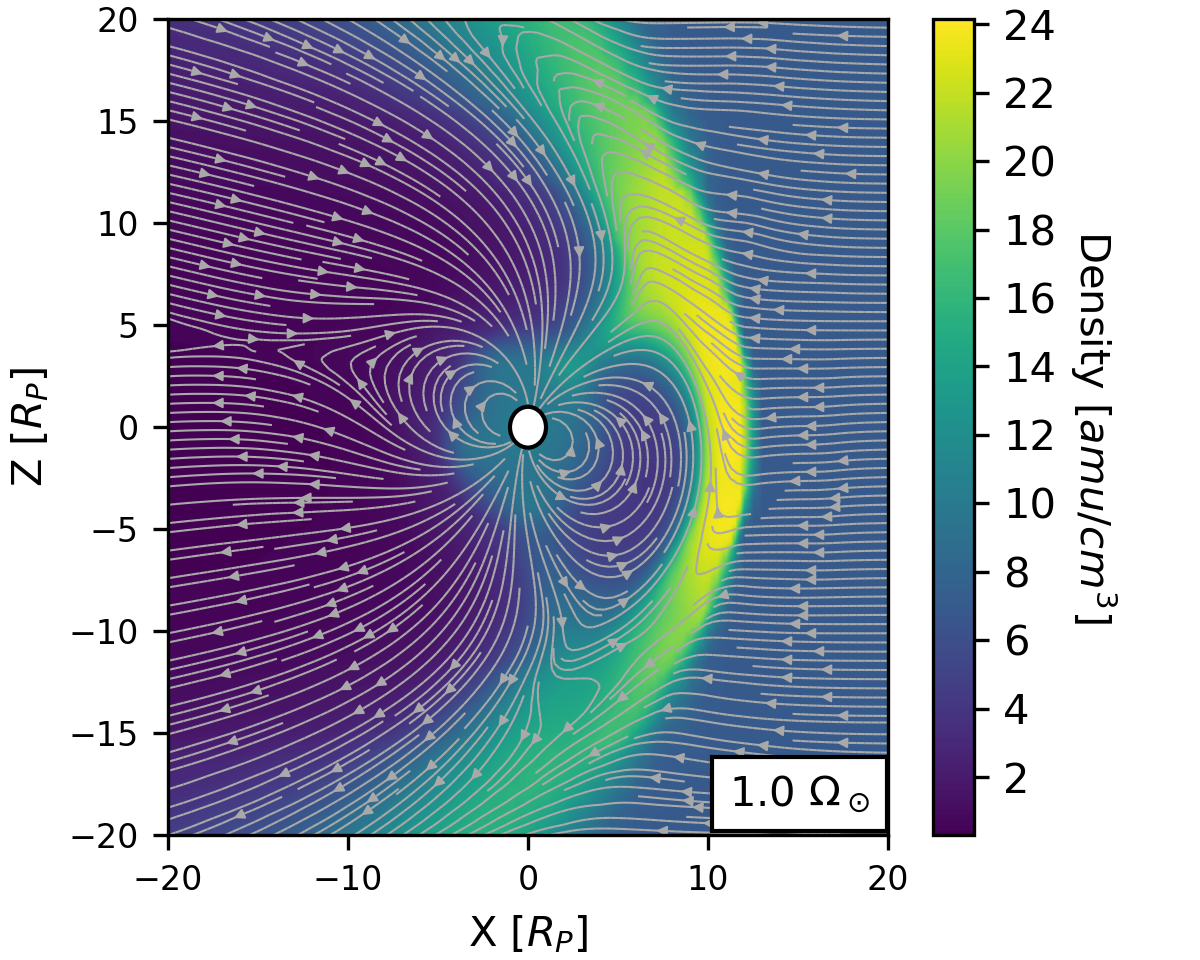}
    \includegraphics[width=\columnwidth]{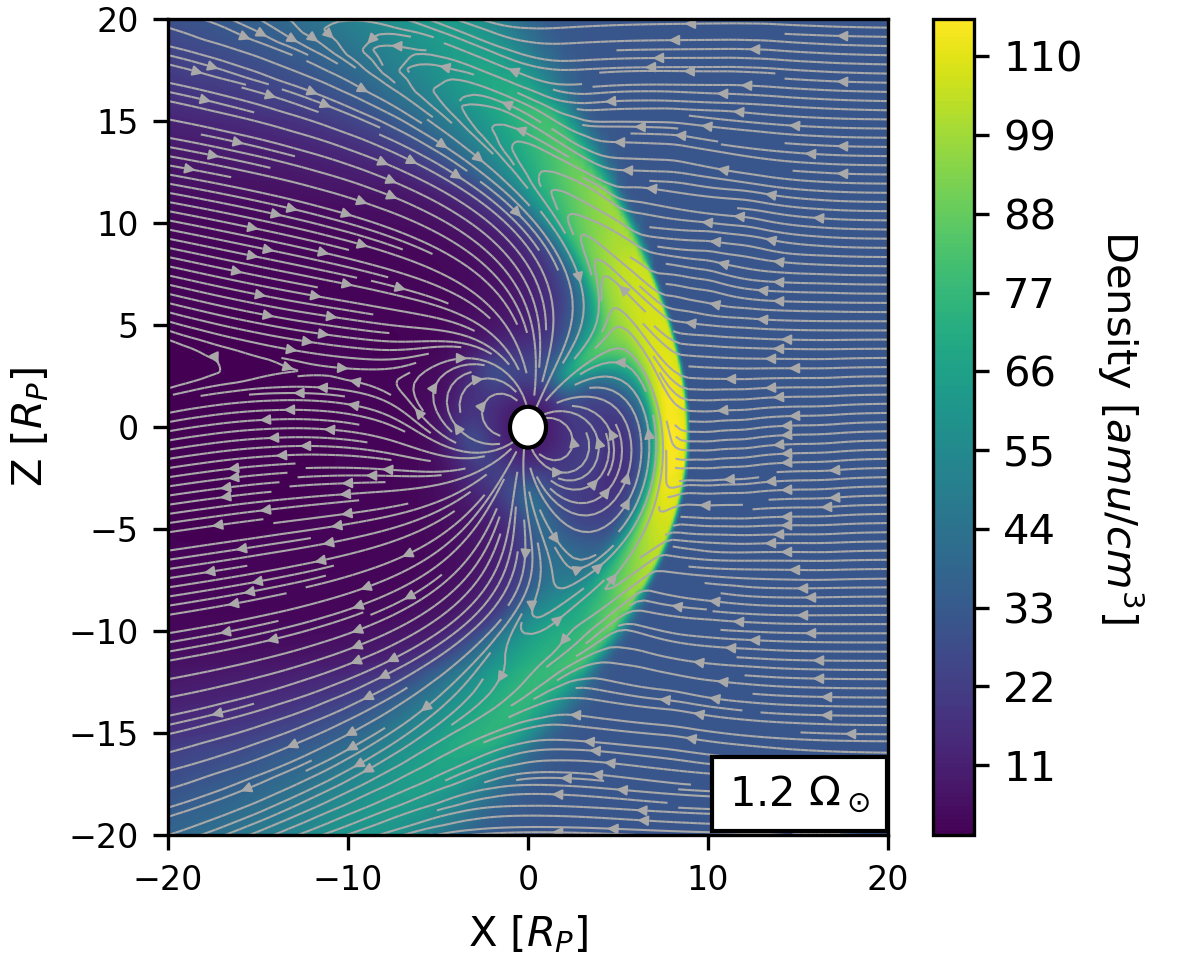}
    \caption{Earth's magnetosphere for different values of stellar rotation from $0.8$~$\Omega_\odot$ to $10$~$\Omega_\odot$. For each model the density distribution in the X-Z plane is shown as a contour. The streamtracers show the magnetic field lines, illustrating the magnetospheres in our models.}
    \label{fig_2d_1}
\end{figure*}

\begin{figure*}
    \includegraphics[width=\columnwidth]{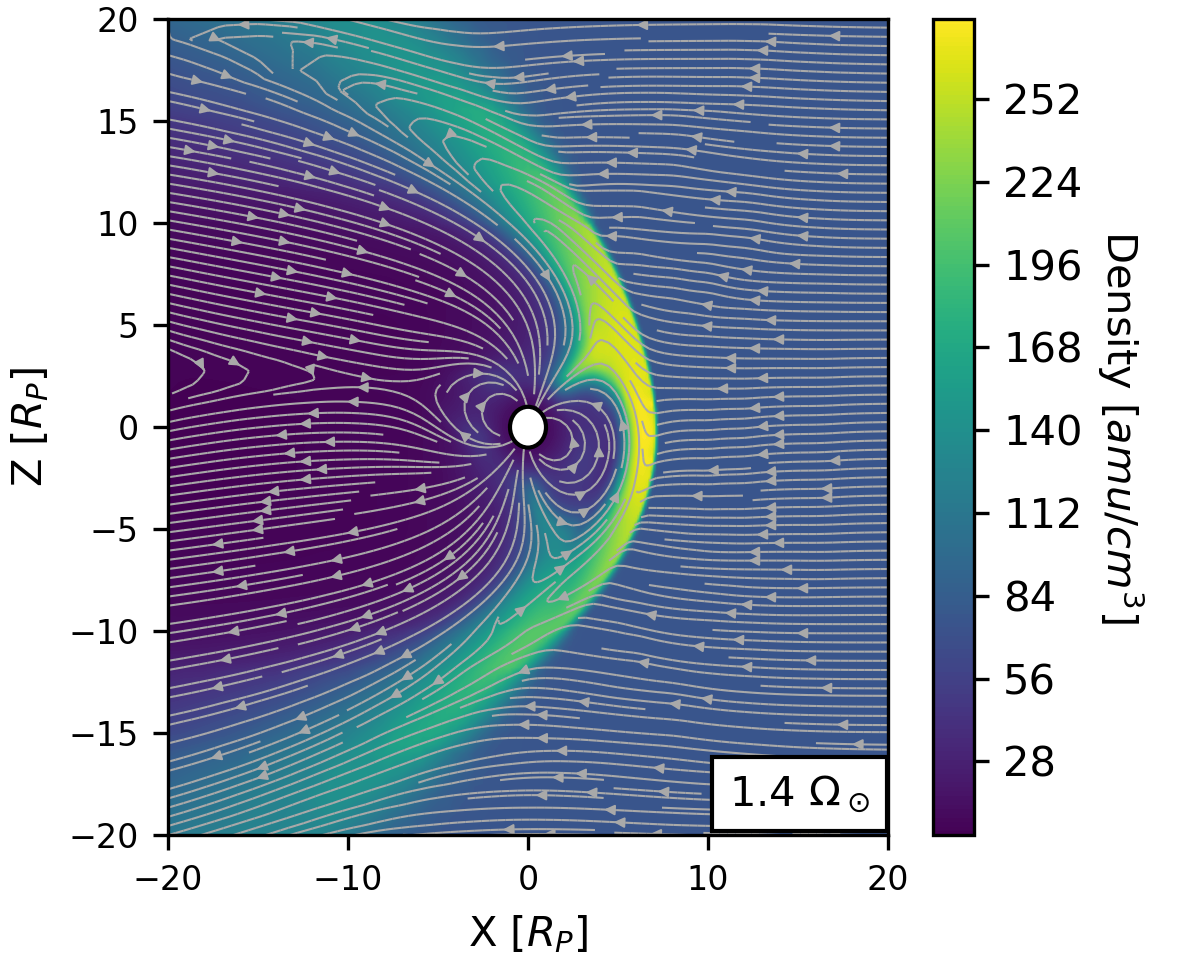}
    \includegraphics[width=\columnwidth]{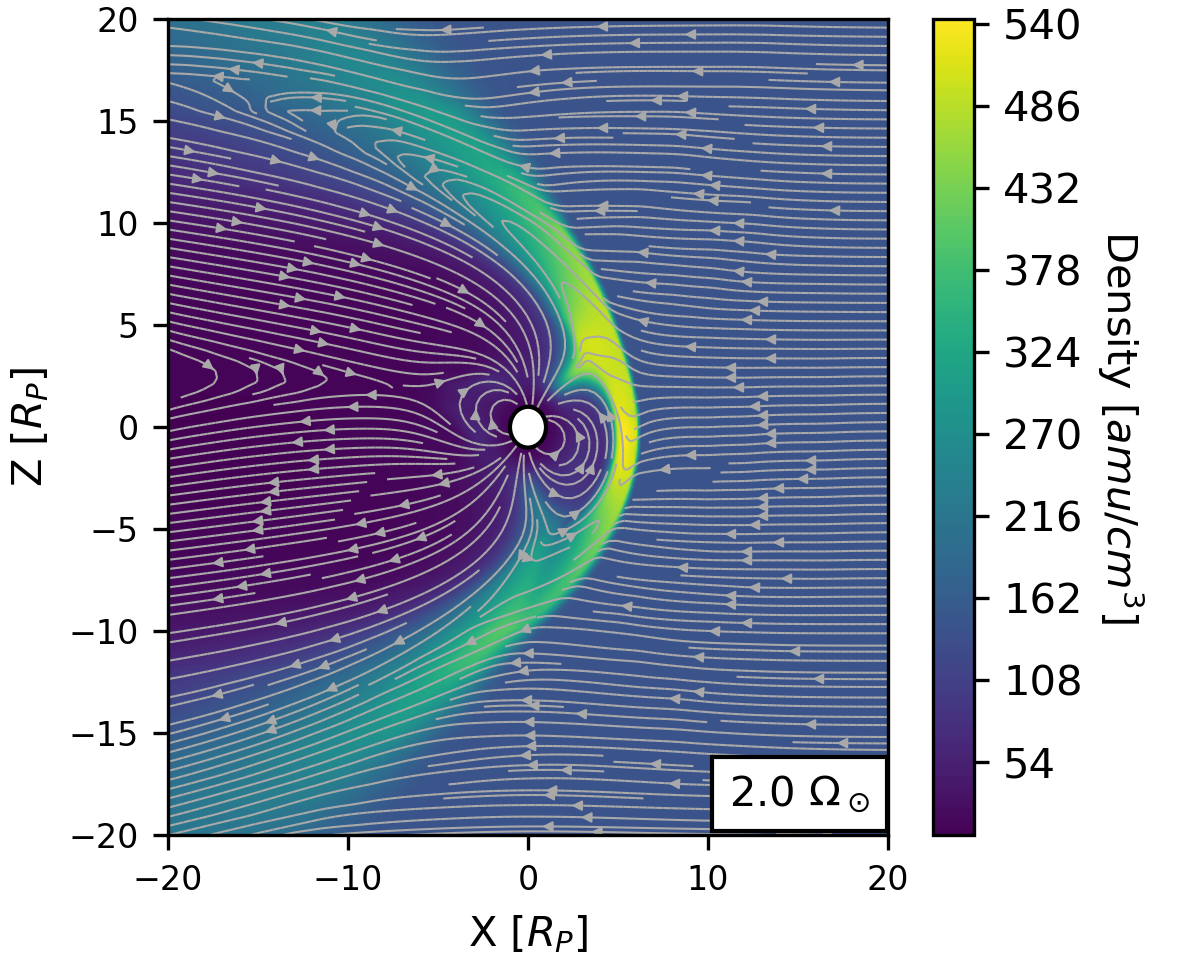}\\
    \includegraphics[width=\columnwidth]{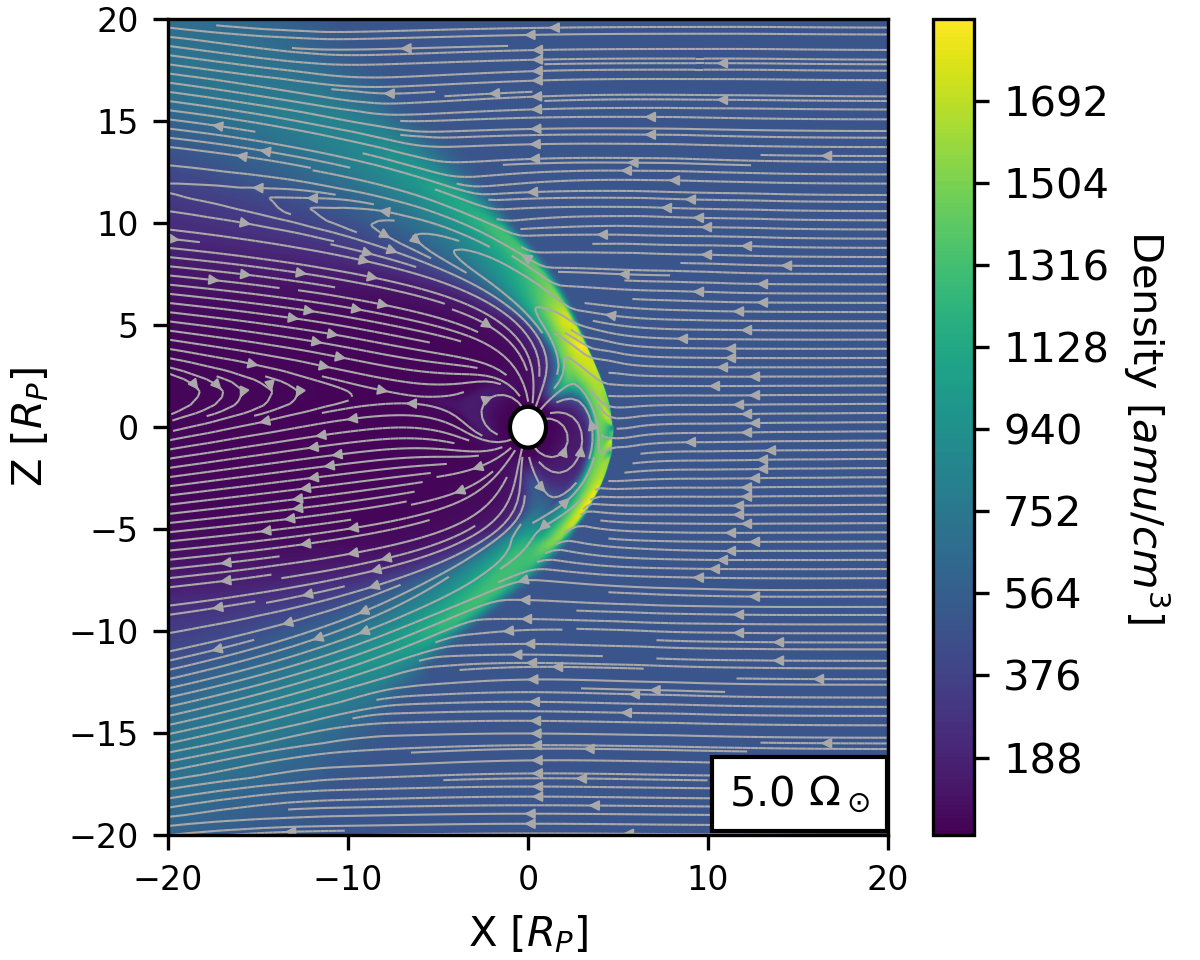}
    \includegraphics[width=\columnwidth]{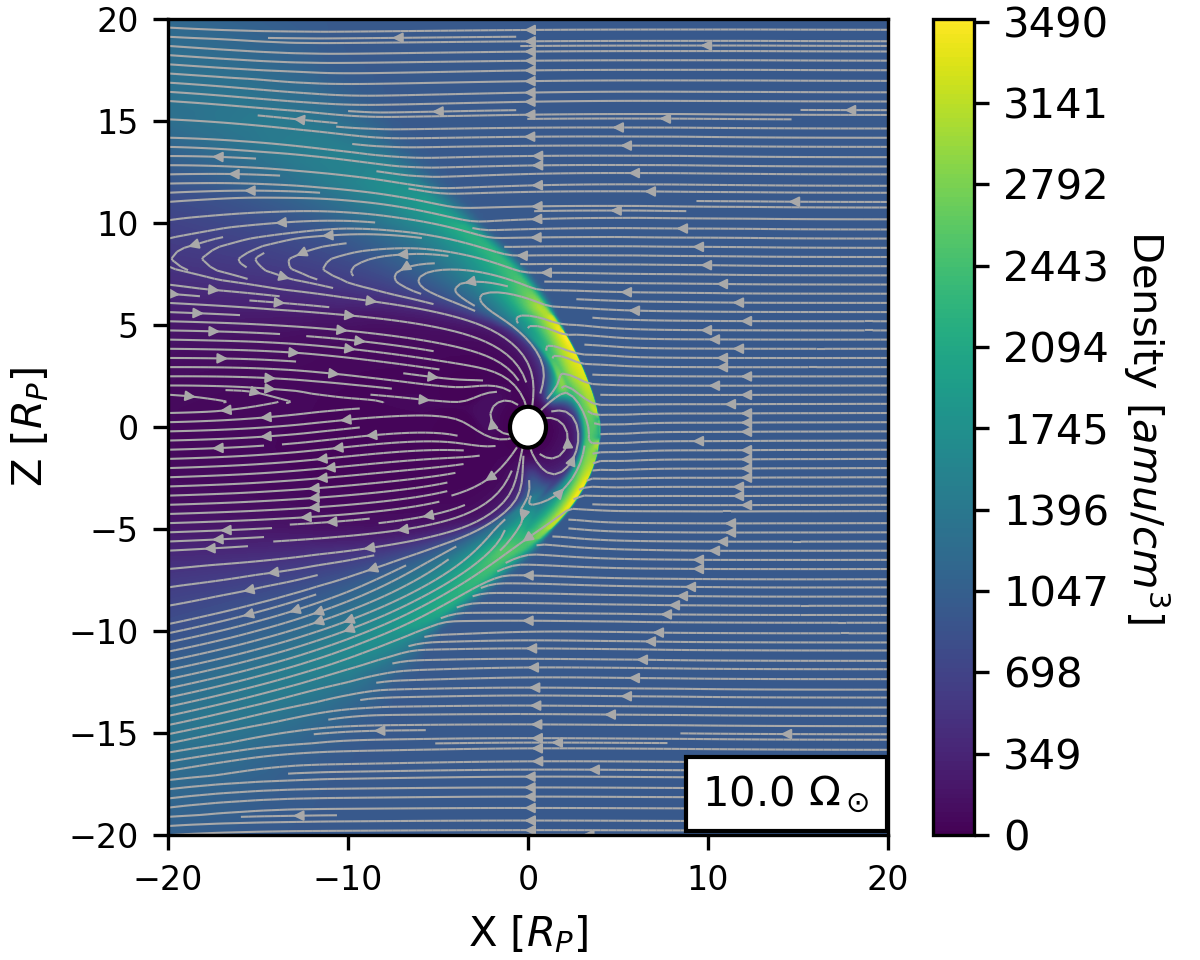}
    \contcaption{} 
    \label{fig_2d_2}
\end{figure*}

Here we examine how the dayside of Earth's magnetosphere varies in time, mainly focusing our analysis along the X axis in each of our 3D models (towards the star).

\subsection{The Magnetopause}
The size of the magnetosphere, the region where the dominant magnetic field is due to the planet, is marked by the magnetopause standoff distance ($r_M$), which analytically can be described as the point at which the magnetic pressure in the plasma due to the planet's magnetic field balances the ram pressure of the stellar wind \citep{Chapman1931}:

\begin{equation}
    r_M^{\rm analyt} = \Bigg( \frac{B_0^2}{8\pi\rho^{\rm sw}(u^{\rm sw})^2} \Bigg)^{\frac{1}{6}} R_p, 
    \label{chapmaneqn}
\end{equation}

\noindent where $B_0$ is the planet's magnetic dipole strength and $\rho^{\rm sw}$ is stellar wind mass density. This equation neglects the presence of the bow shock and magnetosheath, so we will use an alternate method to locate the magnetopause.

\begin{figure*}
	\includegraphics[width=1.0\textwidth]{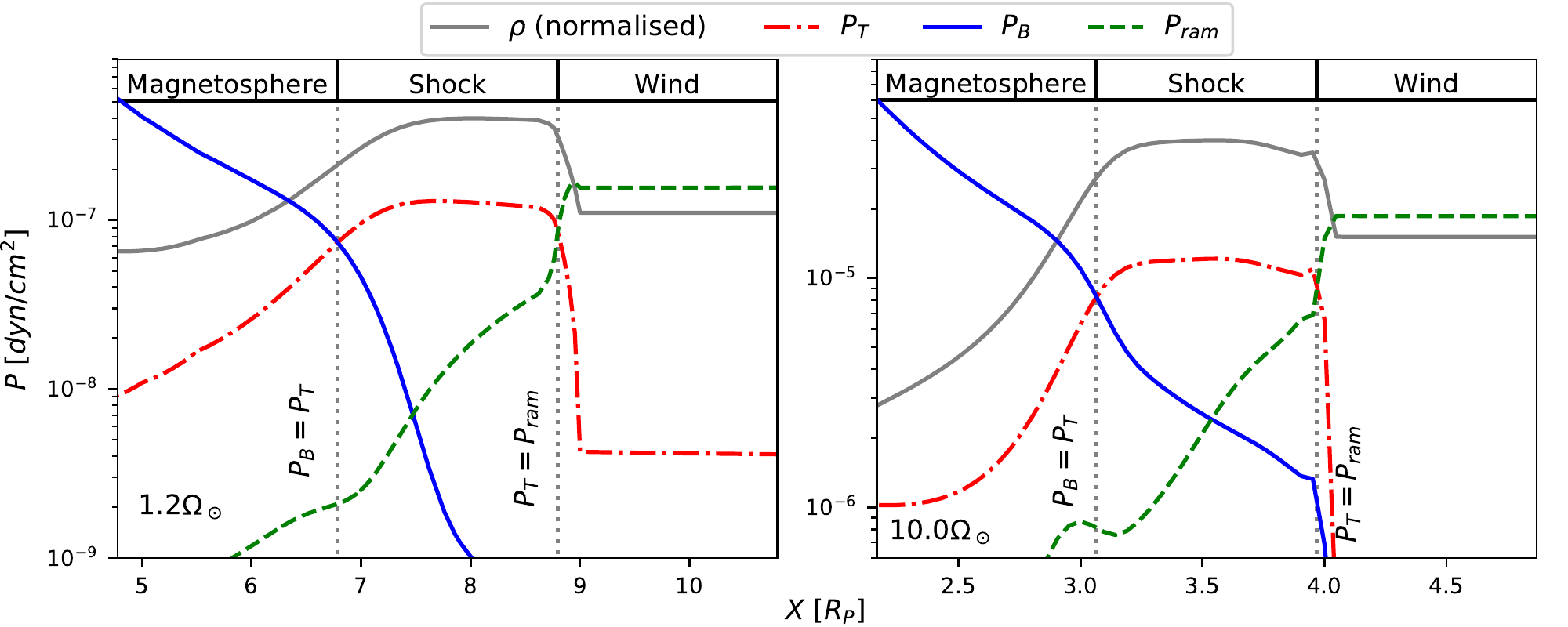}
    \caption{The variation of thermal, magnetic and ram pressures on the day side of the planet towards the star for the $1.2$~$\Omega_\odot$ and $10$~$\Omega_\odot$ models. The grey vertical dotted lines mark the points where thermal-magnetic and thermal-ram pressure balances occur, which represent the magnetopause and bow shock standoff distances. This allows for the size of the magnetosphere and magnetosheath to be established. The solid blue lines are the magnetic pressure, the dot-dashed red are the thermal pressures and the dashed green are the ram pressures.}
    \label{fig_pressure}
\end{figure*}

There are three main contributors to the total pressure in our models: thermal ($P_T = n_pk_BT$); magnetic ($P_B = B^2/8\pi$); and ram pressure ($P_{\rm ram} = \rho u_r^2$). $k_B$ is Boltzmann's constant. To examine which contributor dominates at a certain distance from the planet, these have been separated and plotted alongside the normalized density in Figure \ref{fig_pressure}, for two representative cases: low ($1.2~\Omega_\odot$) and high ($10~\Omega_\odot$) stellar rotation.

For both high and low  stellar rotation rates, we see similar trends in the thermal, magnetic, and ram pressures. Inside the magnetosphere, magnetic pressure dominates due to the strong magnetic field of the planet. In the stellar wind, we see the dominant component is the ram pressure due to the high density, much higher velocity, and relatively low magnetic field and thermal components in the wind. 

When the stellar wind is shocked, the majority of ram pressure is converted to thermal pressure \citep{Cravens2004Book}. In our models, this fraction is 75\%.  In the shock the velocity of the wind drops by a factor of 4 and the density increases by a factor of 4, for models with stellar rotation $<~10~\Omega_\odot$, which exhibit strong shocks.

\begin{figure}
	\includegraphics[width=\columnwidth]{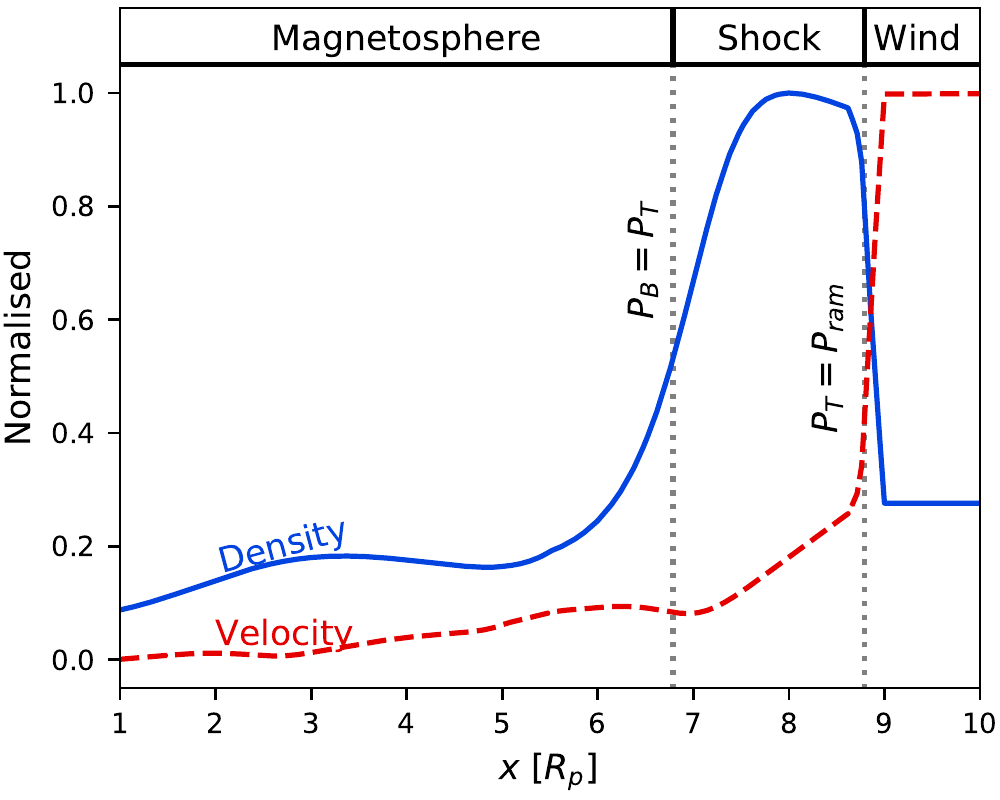}
    \caption{The variation of velocity magnitude and density towards the star in the $1.2$~$\Omega_\odot$ model. These are normalized for comparison. This verifies the pressure balance method for establishing the standoff distance and magnetosheath thickness from our models, as we see the ram-thermal pressure balance correctly identifies the shock location, while the magnetic-thermal balance marks a minimum in velocity. We confirm that these models show adiabatic shocks following the Rankine-Hugoniot conditions, with the density increasing by a factor of 4 and the velocity decreasing by a factor of 4 compared to the stellar wind values.}
    \label{fig_VelAndRho}
\end{figure}

The Chapman-Ferraro equation (Equation \ref{chapmaneqn}) neglects the extent of the shock. It balances the magnetic pressure on the left side (inside magnetosphere) with the ram-pressure on the right side (in the stellar wind). In our simulations we can see that this transition is mediated by the presence of a finite shock (the magnetosheath), dominated by thermal pressure. The magnetopause standoff distance is marked by the point on the X axis where the magnetic and thermal pressures are balanced. Similarly, to identify the bow shock standoff distance we use the balance of thermal and ram pressures. These distances for the $1.2~\Omega_\odot$ model are marked in Figures \ref{fig_pressure} and \ref{fig_VelAndRho}. We can see that the magnetopause corresponds to a local minimum in the velocity magnitude, with a density cavity just inside this. This cavity is carved by the large closed magnetic field lines in this region, which lead shocked material away.

As faster rotating stars are more active, they have stronger stellar winds. Therefore we expect a faster rotator to induce a smaller magnetosphere around the Earth, as the stronger stellar wind leads to a larger ram pressure. This is seen in Figure \ref{fig_standoff_omega}, as we see a gradual decrease in standoff distance with increasing $\Omega$ (Table \ref{table_colatitude}). There is a break in the trend of standoff distance with stellar rotation at approximately $1.4$~$\Omega_\odot$. This is due to how the base temperature of the winds is specified in Section \ref{stellar_wind_section}, which is given by a piece-wise function about $1.4$~$\Omega_\odot$. To find a relation in terms of $\Omega$ we fit our data using a piece-wise function. We find the standoff distance varies with $\Omega$ according to the following relation:

\begin{figure}
	\includegraphics[width=\columnwidth]{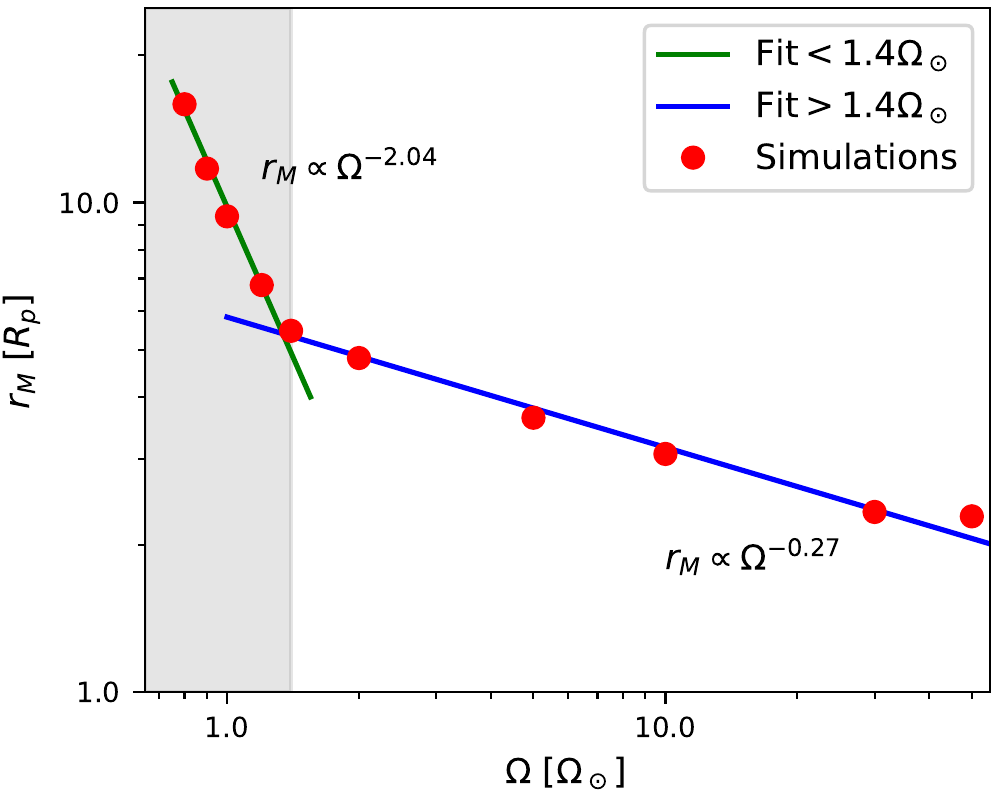}
    \caption{The variation in standoff distance with $\Omega$ for all magnetosphere models. As was seen in the stellar wind models, we see a different trend in $r_M$ with $\Omega$ around $1.4$~$\Omega_\odot$. Our fits are shown with the green line ($<1.4$~$\Omega_\odot$) and blue line ($\geq1.4$~$\Omega_\odot$), with the gray region marking the lower $\Omega$ domain.}
    \label{fig_standoff_omega}
\end{figure}

\begin{equation}
  r_M \propto \begin{cases}
    \Omega^{-2.04} & \Omega < 1.4 \Omega_\odot \\
     \Omega^{-0.27} & \Omega \geq 1.4 \Omega_\odot
  \end{cases}
  .
  \label{piece-wise-function}
\end{equation}
In our models, we adopted a dipolar field strength of the Earth that is constant in time (the same value across all our simulations). However, \citet{Tarduno2010} found that when the Earth was approximately 1 Gyr old, its magnetic dipole strength was lower than it is today, within the range from 0.5 to 0.7 times the present-day value. \citet{Zuluaga2013} modelled the change in planetary dipole moment with time, suggesting that the  magnetic field strength of the Earth averaged over its evolution is approximately 90\% of its current value (i.e., during most of its evolution, the magnetic moment of the Earth has not changed). Given that $r_M$ depends very weakly on the field strength ($r_M \propto B_0^{1/3}$), the evolving dipole moment would contribute a mean change of 3\% to $r_M$ over the magnetosphere's evolution, when compared to models using present day values at all ages. Additionally, as can be seen in our piece-wise-function (Equation \ref{piece-wise-function}), which has the same break seen in the assumption of stellar wind temperature, the variation in magnetospheric size is much more sensitive to the changing stellar wind than what would be expected due to an evolving planetary dipole \citep[see also][]{dualta2018}. 

\begin{table}
    \centering
    \caption{Properties of the magnetopause extracted from our simulations: the size of the magnetopause ($r_M$),  calculated along the $X$ axis, and the colatitude $\Phi$ of the last open field line on the dayside, which is calculated from the rotational axis of the planet ($23.5^\circ$ from the ecliptic).}
    \begin{tabular}{ccc}
        \hline
        $\Omega$ [$\Omega_\odot$] & $r_M [R_p]$ &$\Phi$ [$^\circ$] \\
        \hline
        0.8 & 15.9 & 11.1\\
       0.9 & 11.7 & 10.2\\
       1.0 & 9.4 & 10.3\\
       1.2 & 6.8 & 10.5\\
       1.4 & 5.5 & 11.3\\
       2 & 4.8 & 12.1\\
       5 & 3.6 & 13.4\\
       10 & 3.1 & 17.1\\
       30 & 2.4 & 26.2\\
       50 & 2.3 & 31.8\\
        \hline
    \end{tabular}
    \label{table_colatitude}
\end{table}

As was discussed in Section \ref{sec_intro}, the shape and size of the magnetosphere has important implications on atmospheric loss. The area on the planet connected to open magnetic field lines determines the extent to which stellar wind material can reach the planet (We call this scenario ``impact" from now on). If this area is large, a greater amount of stellar wind will impact the atmosphere, which could induce atmospheric loss. If this area is small, there are a greater number of closed field lines, which can collect stellar wind plasma and focus it onto the atmosphere (here-on referred to as ``collection"), which also could cause the same effect of enhancing atmospheric loss \citep{blackman2018}.

Which of these two competing mechanisms (impact vs collection) causes more harm to the planet over its lifetime is currently still in debate. We do not model atmospheric loss here. However from our models we can examine the area on the planet connected to open field lines, which is quantified by the colatitude $\Phi$ of the last open field line on the dayside of the North pole in the X-Z plane. These are given in Table \ref{table_colatitude}.

In the young system, the area covered by open field lines is much larger than for lower stellar rotation rates. This increased area could lead to a higher rate of stellar wind impact than in an old system. With the stellar wind gradually relaxing, the difference between young and old systems is further enhanced. As the magnetosphere is small in the earlier stages of the Earth's evolution and the area of open field lines is large, we expect that stellar wind impact will dominate over plasma collection in the young systems.

As the Sun spun down, this area gradually decreased, with a colatitude of open field lines of approximately $10.3^\circ$ obtained for present day. Analytically, the colatitude of open field lines can be found from the magnetopause standoff distance through the following expression \citep{Vidotto2013}:

\begin{equation}
    \Phi^{\rm analyt} = \arcsin\Bigg(\sqrt{\frac{R_p}{r_M}}\Bigg).
\end{equation}

\noindent Using this expression, we find $\Phi =18.8^{\circ}$ for the present day, using $r_M$ from our models. This value is significantly higher than what we obtain from our simulation. We attribute this difference to the orientation of the magnetic fields in the system. For example, in the case of magnetic fields are anti-parallel on the dayside, which we examine in Appendix \ref{Zcomp}, the numerical and analytical values agree with each other.

With the magnetosphere gradually growing in size, the stellar wind gradually relaxing and the area of open field lines decreasing, it is clear that the amount of stellar wind inflow will decrease with time. Contrary to this, the collecting area for stellar wind plasma will increase through Earth's evolution, as the magnetospheric size and area covered by closed field lines increases. It is clear that of the two competing effects, stellar wind inflow will dominate in the young system, whilst plasma collection will dominate at old ages. 

\subsection{The Bow Shock and Magnetosheath}

The bow shock standoff distance is the distance from the centre of the planet towards the star where the shock wave is formed. In line with how we define the magnetosphere standoff distance, we use the balance of thermal and ram pressures to establish the bow shock standoff distance. At this point a shock wave is formed as the supermagnetosonic stellar wind encounters the magnetic field of the planet. The region between this point and the magnetosphere standoff distance contains the shocked material and is known as the magnetosheath. As outlined by \citet{Balogh2013, Gombosi2004, Spreiter1966}, the bow shock distance ($r_{BS}$) is related to the magnetopause standoff distance according to the following relation:

\begin{equation}
    r_{BS}-r_M = \Delta r \approx 1.1 \frac{N_2}{N_1} r_M,
    \label{thick_eqn}
\end{equation}

\noindent where the density compression factor $\frac{N_2}{N_1}$ (ratio of the density in front of the shock, and that in the shock) is given by the inverse of the equation below

\begin{equation}
    \frac{N_1}{N_2} \approx \frac{\gamma +1}{\gamma -1} - 2 \frac{\gamma+1}{\mathcal{M}^2(\gamma-1)^2},
    \label{compression_ratio}
\end{equation}

\noindent where $\gamma=5/3$ is the adiabatic index. In a strong shock this ratio is approximately 4, as the Mach number is large and the second term may be neglected.

The strength of the shock is determined by the Mach number of the stellar wind (see Table \ref{wind-table}). If this is much greater than 1, a strong shock is formed and so the obtained magnetosheath thickness should be $0.275$~$r_M$ \citep{Gombosi2004} in the strong shock models ($\leq~10~\Omega_\odot$). Using the above relation we can predict the expected bow shock distance and compare this to what we obtain in our simulations. 

\begin{figure}
	\includegraphics[width=\columnwidth]{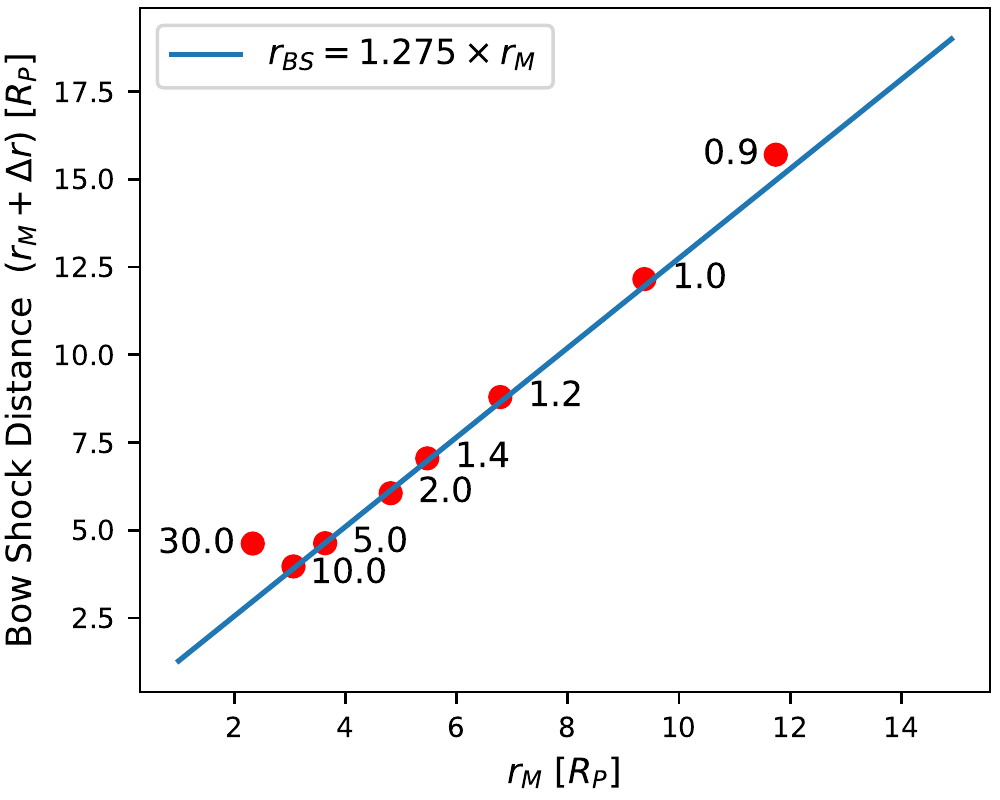}
    \caption{The obtained bow shock distance (standoff distance $+$ magnetosheath thickness) vs each magnetosphere standoff distance. We see the results from our models follow closely Equation \ref{thick_eqn} for a strong shock $\mathcal{M} \gg 1$.}
    \label{fig_BS_vs_Stand}
\end{figure}

The relationship between bow shock distance and standoff distance can be rationalized by considering the pressures and forces on the shocked material. On the planet-side of the planet-star line, the magnetic field of the planet is exerting a force on the shocked material directed away from the planet. On the star side, the wind is exerting a force directed in the opposite direction, as wind particles impact shocked material are themselves shocked. For a smaller standoff distance, the stellar wind must be relatively stronger. As a result, both the ram pressure and the magnetic pressure will be stronger as the shock is closer to the planet. Therefore the net force acting on the bow shock will act to reduce the thickness of the shock for a lower standoff distance. For a weaker wind the opposite effect is achieved. 

In our models, a strong correlation between the bow shock distance and the prescribed relation in Equation (\ref{thick_eqn}) for the strong shock ($\mathcal{M}\gg1$) in the $<10$~$\Omega_\odot$ models. This confirms that as the host star spins down during the main sequence the magnetosphere and magnetosheath both increase in size proportional to the piece-wise function in Equation (\ref{piece-wise-function}) for the majority of its evolution. The $30~\Omega_\odot$ model deviates from this trend. This is due to the weaker shock exhibited in this model, which is discussed in Section \ref{sec_high_omega}.

\begin{figure}
	\includegraphics[width=\columnwidth]{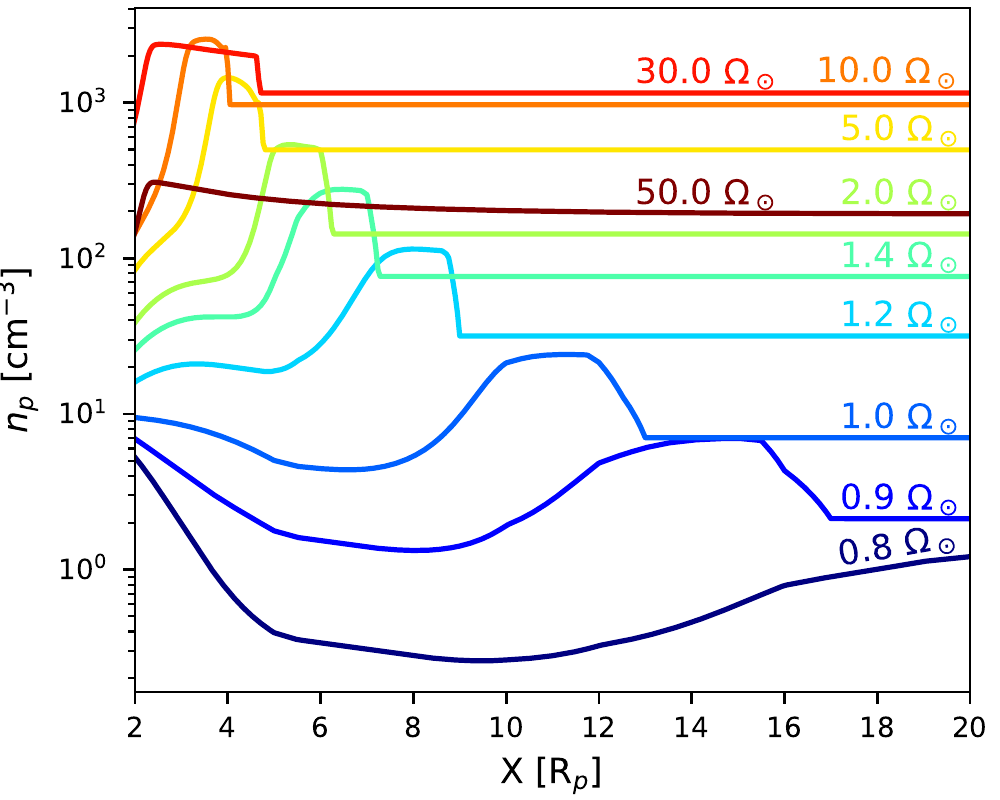}
    \caption{Variation of density on the dayside of the planet, towards the star (X axis) for different values of stellar rotation $\Omega$. This shows the relative change of density through each of our models, as well as illustrating the Rankine-Hugoniot shock conditions followed in our models (increase by factor of 4 in density).}
    \label{fig_density}
\end{figure}

Figures \ref{fig_2d_1} and \ref{fig_standoff_omega} show that the magnetosphere standoff distance moves closer to the planet for higher stellar rotation (stronger winds). As well as this we see a general decrease in thickness of the bow shock for higher values of $\Omega$, as seen in Figure \ref{fig_BS_vs_Stand}, and an increase in peak density up to $30~\Omega_\odot$ as seen in Figure \ref{fig_density}. In this figure, we see the density distribution is constant at large X for different stellar rotation rates. This indicates the extent to which the stellar wind penetrates into our computational grid. There is a jump in density closer to the planet, signifying the formation of a bow shock. For most of our models these shocks follow the Rankine-Hugoniot jump conditions, which predict for an adiabatic shock a factor of 4 increase in density. The exceptions to this are the $30~\Omega_\odot$ and $50~\Omega_\odot$ models, showing a weaker and no shock respectively. This is discussed in Section \ref{sec_high_omega}.

\begin{figure*}
    \includegraphics[width=\textwidth]{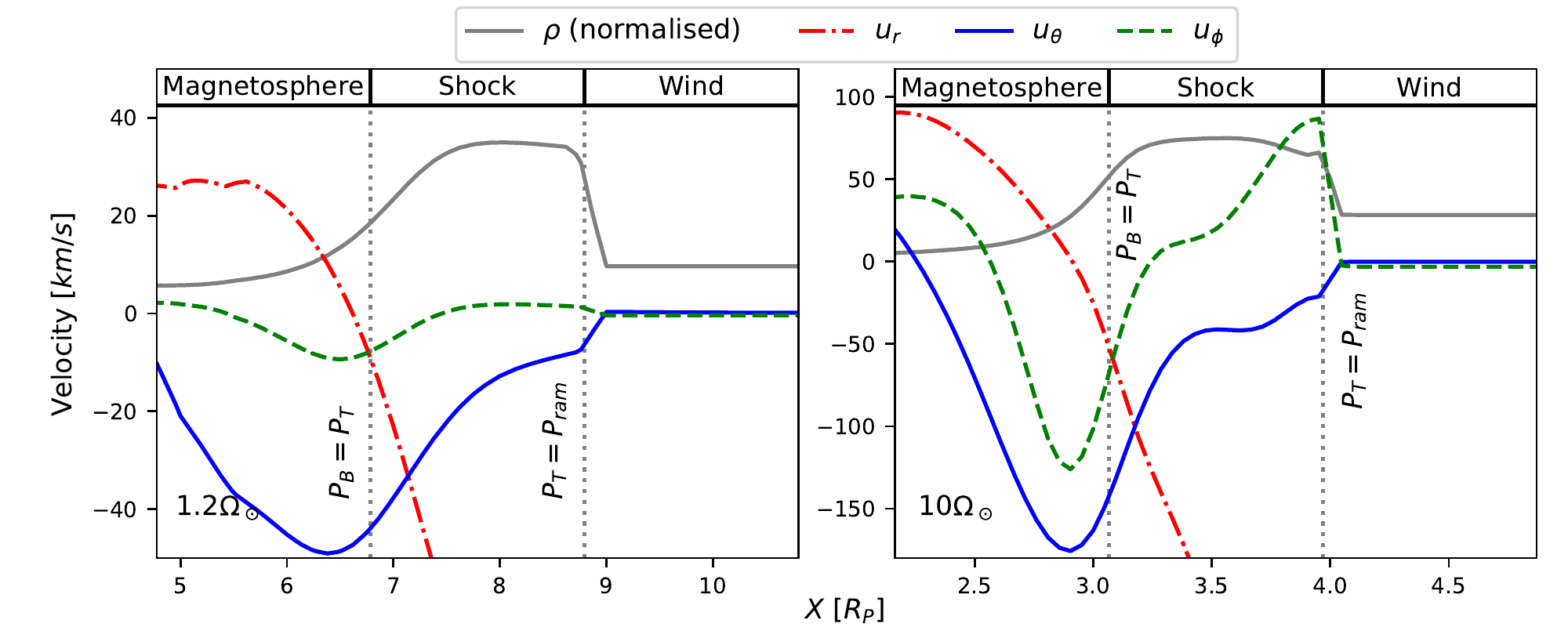}
	\includegraphics[width=\textwidth]{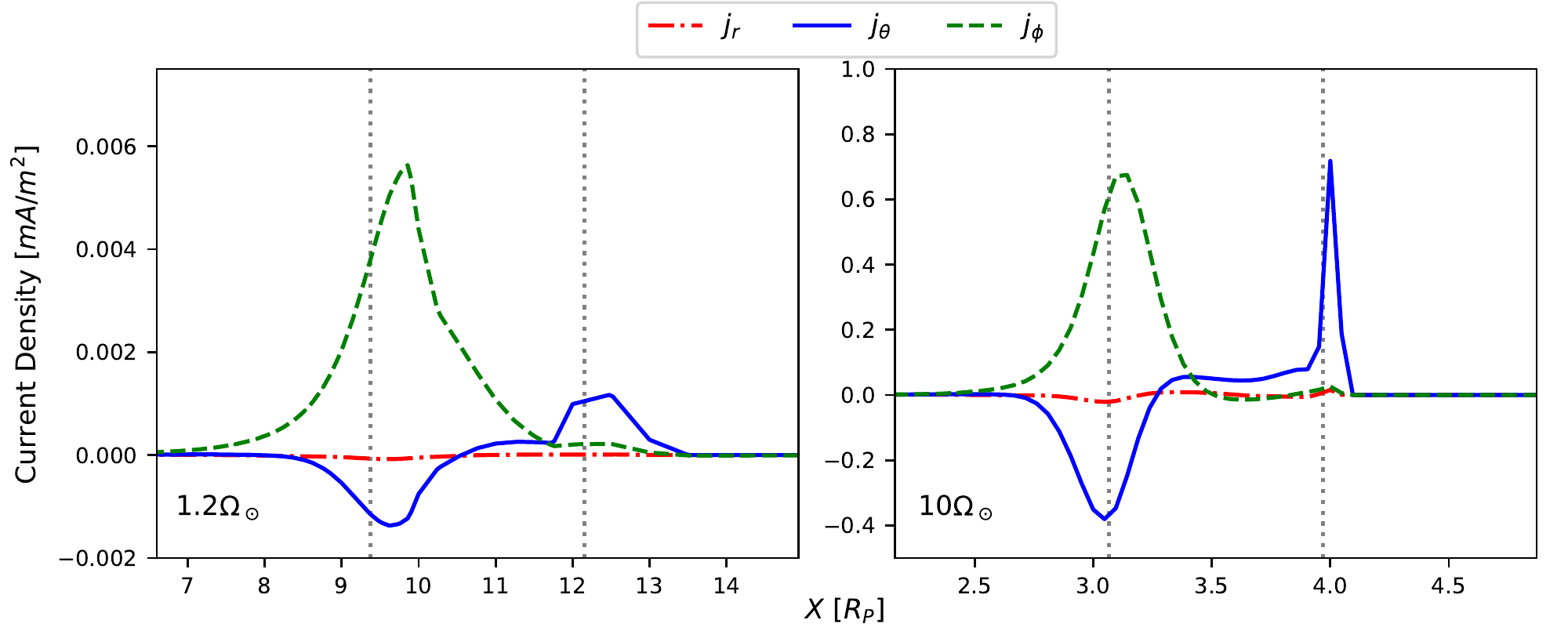}
	\includegraphics[width=\textwidth]{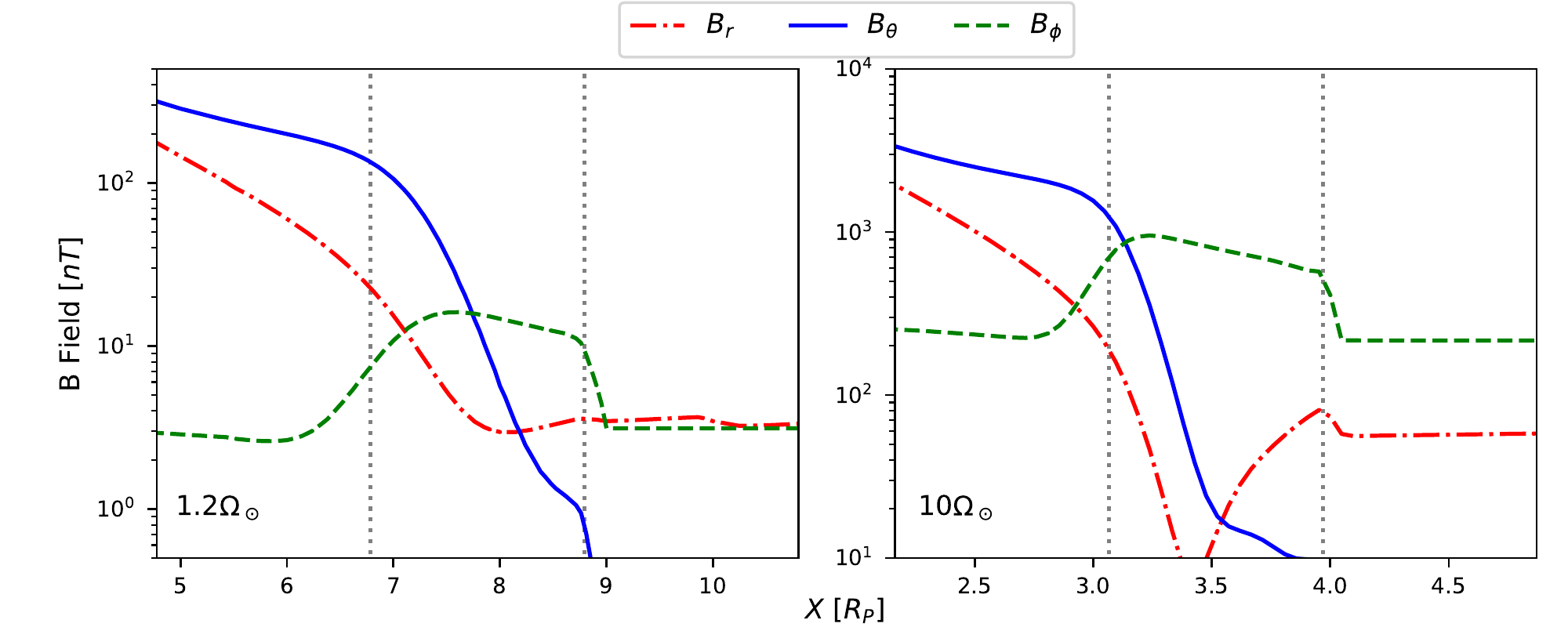}
    \caption{The variation velocity, current and B-field components along the x axis towards the star, for the $1.2~\Omega_\odot$ and $10.0~\Omega_\odot$ model. The grey vertical lines again mark the point where magnetic-thermal (left) and ram-thermal (right) pressures are balanced. The red dot-dashed line are the radial components, the solid blue are the $\theta$ components, and the dashed green are the $\phi$ components.}
    \label{fig_velocities}
\end{figure*}

In the case of large magnetospheres (lower stellar rotation) we see a low density cavity within the magnetosphere. To understand why this occurs we compare the density and velocity magnitude in the model, as seen in Figure \ref{fig_VelAndRho}. This minimum in density occurs at the point where the velocity magnitude is at a local maximum. This region corresponds to where some of the largest closed magnetic field lines from the planet exist, seen in Figure \ref{fig_2d_1}. The top panels of Figure \ref{fig_velocities} shows the dominant components of velocity in this region are positive $u_r$ and negative $u_\theta$. (In the higher $\Omega$ models, $u_\phi$ begins to become significant). On the boundary of the magnetopause, the $u_\theta$ component is at a maximum magnitude. Along the X axis this corresponds to the Z direction, which is perpendicular to the orbital plane of the planet in this instance. The high $u_\theta$ velocity here shows that the magnetic field lines lead a significant amount of material away. Just inside this region, the positive $u_r$ component is moving material away from the planet and towards the sweeping $\theta$ component. As a result there is an under-density in this region. Figure \ref{fig_velocities} shows that the $u_\phi$ component is relatively negligible along the subsolar line. Finally outside the bow shock we can see $u_r$ is at a maximum, which is to be expected as the stellar wind is mostly radial.

Across the shock, we see little variation in the radial component of the magnetic field (Figure \ref{fig_velocities}, bottom panel). In the $B_\phi$ component however we see a small shock like behaviour in both low and high stellar rotation models, with an overall increase of a factor of 4 (adiabatic shock).

The largest variation in the magnetic field is observed in $B_\theta$. Just inside the magnetosphere, this component is strong and positive, corresponding to a northward (positive Z) direction of the dipolar field. This corresponds to the largest closed field lines in the X-Z plane, which at this point on the X axis will be oriented in the positive Z direction. This is true for all models. Through the bow shock, there is a steep decrease in this $B_\theta$ component for both high and low rotation models. This illustrates how the magnetic field lines across the magnetopause transition from closed to open field lines in the X-Y plane.

The middle panel in Figure \ref{fig_velocities} shows two notable trends in current, which are similar in both low and high $\Omega$ models. The $j_\phi$ component shows the ``Chapman-Ferraro" current, also known as the magnetopause current. This current separates the shocked magnetosheath from the relatively empty magnetopause. This prevents the terrestrial dipole field from penetrating into the solar wind \citep{Gombosi2004}. At the boundary between the magnetopause and bow shock, there is a strong negative $j_\theta$ component. At the boundary between the stellar wind and bow shock, the $j_\theta$ component is strong and positive. This can be explained by the variation in $B_\phi$. This tangential component to the shock surface itself shows shock-like behaviour, transitioning from low values in the magnetosphere, to shocked values in the magnetosheath before returning to low in the stellar wind. This variation in $B_\phi$ generates a perpendicular current component, with a negative $j_\theta$ generated by the difference the magnetic field orientation and strength on opposite side of the magnetopause ($B_\theta$ dominated inside, $B_\phi$ dominated outside), and a positive $j_\theta$ when it decreases across the shock.

\section{Very fast rotating young Sun?}\label{sec_high_omega}
 It is not currently known the rotation rate of the Sun at early ages. Recently, a neat study based on the composition of volatile elements in lunar samples from \citet{Saxena2019} suggested that the Sun would have been a slow rotator -- these authors argue that slow rotators are expected to generate lesser amount of energetic particles through coronal mass ejections (CME) and thus would not completely deplete  the sodium and potassium still seen in the moon regolith. To reach such a conclusion, the authors rely on a correlation between solar flares and CMEs \citep[e.g.,][]{2011SoPh..268..195A}, and extend it to younger Sun-like stars using observed flare frequencies from solar analogues. As a consequence, a slowly rotating (less active) young Sun would present less frequent CMEs than a fast rotating young Sun. However, there has been some suggestions in the literature that the solar CME-flare relation might actually over-predict the number of CMEs in younger and more active stars \citep{2013ApJ...764..170D}. In this case,  a fast rotating Sun could actually have had a lower frequency of CMEs (or more `failed CMEs', see \citealt{2018ApJ...862...93A}) for its predicted flare rate at young ages. This could have an impact in the calculated sputtering of volatile elements recorded in lunar data sample. Here, we speculate what would have happened to the Earth's magnetosphere in the case that the young Sun was a  fast rotator.
 
\begin{figure*}
    \includegraphics[width=\columnwidth]{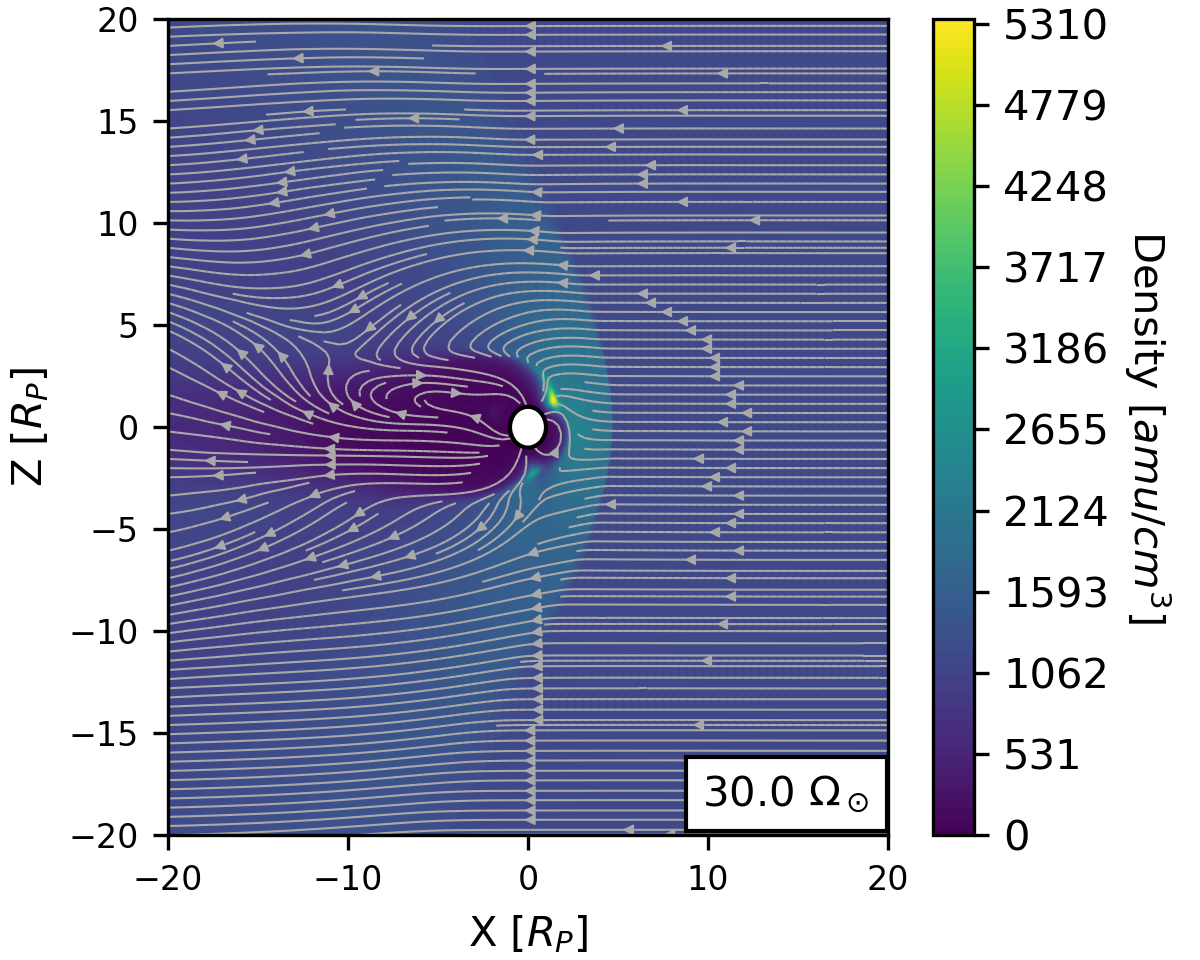}
    \includegraphics[width=\columnwidth]{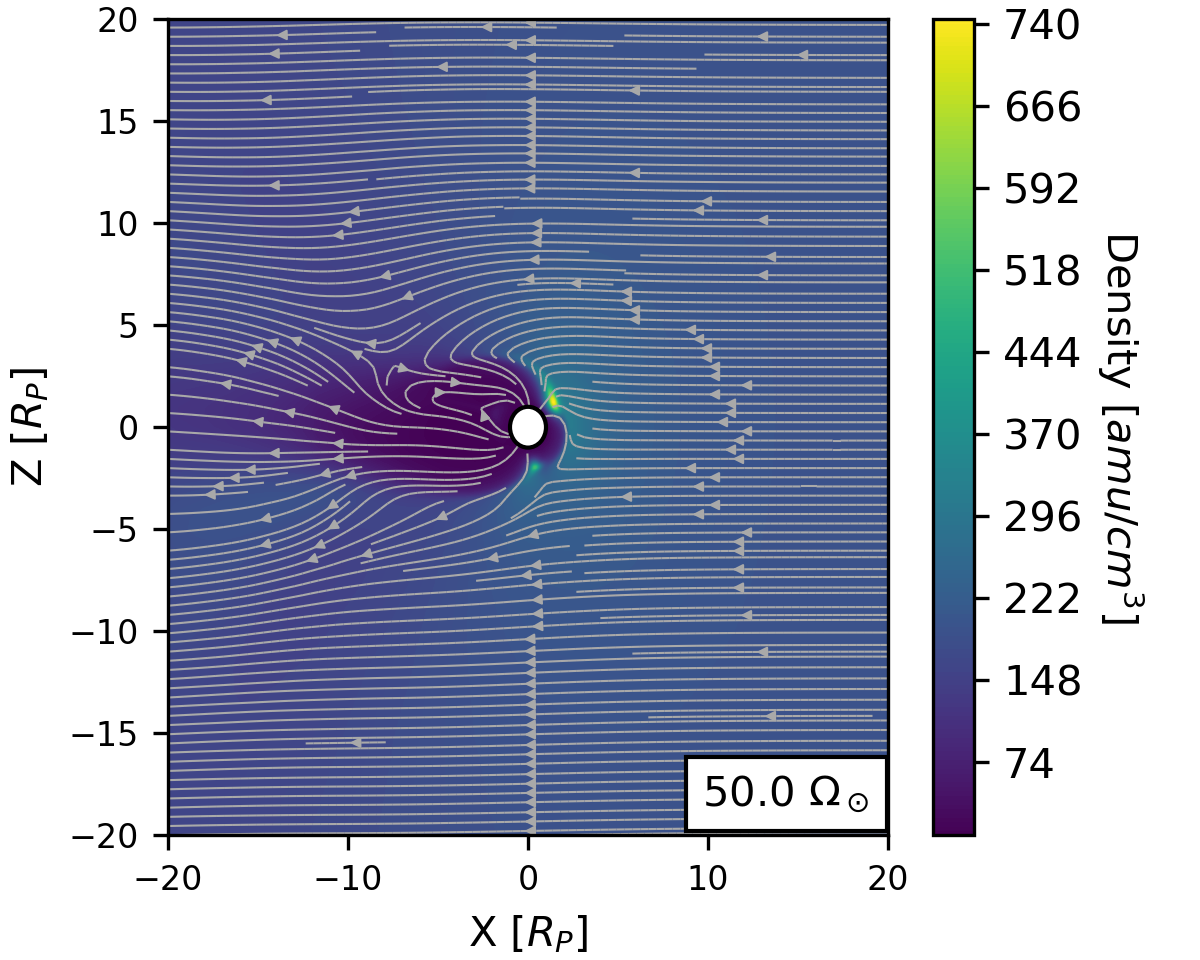}
    \caption{Earth's magnetosphere in the wind of a very fast rotating young Sun. In the $30~\Omega_\odot$ model (left) we see a weaker shock than those in Figure \ref{fig_2d_1}, with a comparably inflated magnetosheath. In the $50~\Omega_\odot$ model (right) there is no shock present, as the wind is submagnetosonic. Despite this, the magnetosphere is not completely crushed, though is reduced to a magnetopause standoff distance of $\approx 2.3~R_p$.}
    \label{fig_2d_high_omega}
\end{figure*}

 In the fast rotating regime it is possible for the young Sun to have a rotation rate as high as $100~\Omega_\odot$, while models of slow rotators suggest a maximum of approximately $5~\Omega_\odot$ \citep[][see also our Figure \ref{fig_gallet}]{Gallet2013}. To understand the magnetosphere in the fast young system, we perform two additional simulations for stellar rotation rates of $30~\Omega_\odot$ and $50~\Omega_\odot$. The results of these models can be seen in Figure \ref{fig_2d_high_omega}.  

\begin{figure*}
    \includegraphics[width=\textwidth]{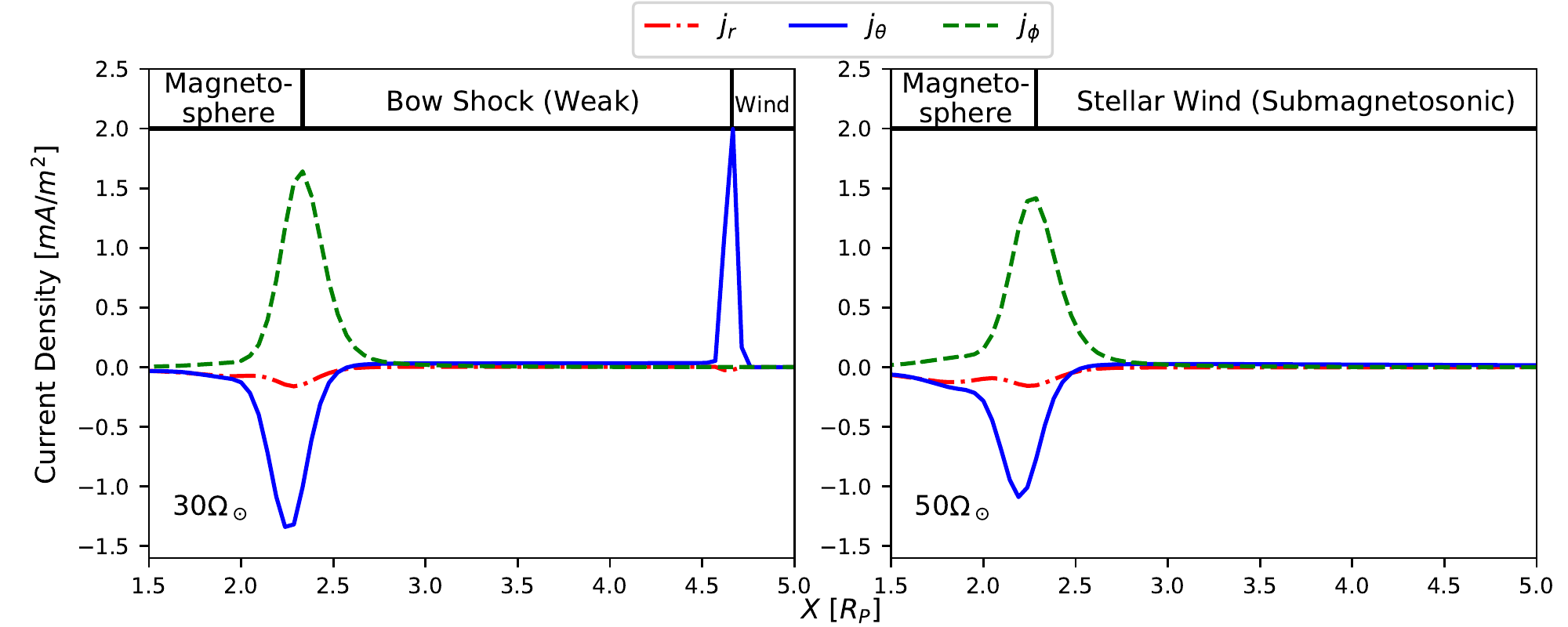}
    \caption{The current components along the subsolar line in the $30~\Omega_\odot$ (left) and $50~\Omega_\odot$ (right) models. The positive $j_\phi$ component is the magnetopause current, which is a current system flowing around the magnetosphere. The positive $j_\theta$ marks the position of the bow shock, and is generated from the change in magnetic field in the shock. As there is no shock in the $50.0~\Omega_\odot$, this current density is not seen here.}
    \label{fig_current_high_omega}
\end{figure*}

In both of these models, the thermal pressure does not become large enough within the magnetosheath to balance either the magnetic or ram pressures, due to a weak shock for $30~\Omega_\odot$, and lack of shock for $50~\Omega_\odot$. We therefore cannot use the points of pressure balance to identify the magnetopause and bow shock standoff distances. Instead we use the $j_\phi$ current density as seen in Figure \ref{fig_current_high_omega}. In all our models, there is a positive $j_\phi$ component around the magnetopause. This is the Chapman-Ferraro or magnetopause current mentioned previously in this section. We use this current density to identify the position of the magnetopause in these fast rotating models. To identify the position of the bow shock in the $30~\Omega_\odot$ model, we use the point where ram pressure begins to dominate all other components.

Due to the lower magnetosonic Mach numbers (1.5 and 0.99 respectively), both these models exhibit some differences to those in Figure \ref{fig_2d_1}. In the $30~\Omega_\odot$ model there is a much weaker shock than for lower rotation rates. As a result the density jumps by only a factor of $\approx 2$ in the bow shock, whereas in the stronger shocks this factor was $4$. Due to the larger $N_1/N_2$ compression ratio, resultant from the lower Mach number, the magnetosheath in this model is thicker than expected for a strong shock. This model therefore deviates from the expected Equation for bow shock standoff distance shown in Figure \ref{fig_BS_vs_Stand}.

In the $50~\Omega_\odot$ model, the wind is sub-magnetosonic so no bow shock is formed. However the magnetosphere still exists in this model, marked by the magnetopause current, and is not completely crushed. With the absence of a bow shock in this system, the magnetosheath thickness can no longer be quantified. These models suggest that if the Sun was a fast rotator, the young Earth would first have no bow shock. As the Sun spun down a weak shock would then form, followed by a strong shock for the remainder of the main sequence. The magnetopause would have had a minimum value of approximately $2.3~R_p$ in this system. For an intermediate or slow rotating system, the Earth would have been surrounded by a strong shock for the entirety of its evolution.

\section{Conclusions}
\label{sec_conclusions}
In this work, we study the evolution of the Earth's magnetosphere over the main-sequence lifetime of our Sun. The novelty of our work is that we coupled two sets of simulations: one to model the evolution of our solar wind and the other to model the evolution of Earth's magnetosphere. The results obtained in the former set of simulations were used as external boundary condition for the second set of simulations. We simulated the evolution of the solar wind using 1.5D stellar wind models of Sun-like stars with different stellar rotation rates. We simulated these winds out to 1 au. We see a split in stellar wind properties at 1 au around $1.4 \Omega_\odot$, in line with how we specify relations for base temperature.  The resulting temperature, density, velocity and magnetic field vectors at 1au were then employed in our 3D magnetosphere models.

Using the balance of thermal and magnetic pressures, we found that the magnetopause standoff distance varies according to the following piece-wise function due to the break in stellar wind properties: $r_M (< 1.4\Omega_\odot) \propto \Omega^{-2.04}$ and $r_M (\geq 1.4\Omega_\odot) \propto \Omega^{-0.27}$. This suggests that given the early solar wind strength, the young Earth's magnetosphere was much smaller than it is today. As the Sun spun down, this size gradually increased, before experiencing greater a increase once the Sun's rotation rate dropped to $1.4 \Omega_\odot$. Furthermore we can predict a much larger magnetosphere size in the future, according to the steep increase with magnetopause distance with decreasing $\Omega$ for $<1.4\Omega_\odot$. Our models yield a standoff distance of $9.4~R_p$ for the present day magnetosphere, which is within the bounds of observed values.

We found a linear relationship between magnetopause standoff distance and the thickness of the magnetosheath for stellar rotations $\leq10~\Omega_\odot$. This is in line with the relation prescribed by \citet{Gombosi2004} and \citet{Balogh2013}, for example, for a strong shock ($\mathcal{M}\gg1$). These stellar wind models all have a $\mathcal{M}\gg1$, allowing us to use Equation (\ref{thick_eqn}) to predict that this thickness will be approximately $0.275$~$r_M$ for a given magnetopause distance (Figure \ref{fig_BS_vs_Stand}). Therefore, we can say that the magnetosheath thickness is proportional to stellar rotation rate in the same way as the magnetopause distance for the majority of its evolution, according to the following piece-wise function: $\Delta r (< 1.4\Omega_\odot) \propto \Omega^{-2.04}$ and $\Delta r (\geq 1.4\Omega_\odot) \propto \Omega^{-0.27}$.

We examined the variation of parameters along the subsolar line (line from the centre of the planet towards the star). We see that in our models the magnetosheath is dominated by thermal pressure, while ram pressure dominates in the stellar wind and magnetic pressure dominates in the magnetosphere. We use the balance of magnetic and thermal pressures to define the magnetopause standoff distance, and the balance of thermal and ram pressures to define the boundary between the magnetosheath and the stellar wind. In our models, current densities also mark both of these positions well. We see strong $\phi$ and $\theta$ currents at the magnetopause boundary, corresponding to the ``Chapman-Ferraro" current, with the $\theta$ component being strong and positive at the boundary of the stellar wind, due to the varying $\phi$ magnetic field at this boundary.

We examined the colatitude if the last open magnetic field line, to discuss whether stellar wind inflow or plasma collection poses a greater threat to Earth's atmosphere at a certain stage of its evolution, which is currently in debate \citep{blackman2018}. We found that this colatitude decreases through the evolution of Earth's magnetosphere. This accompanied by the gradual increase in the size of the magnetosphere suggests that stellar wind inflow would pose the greatest threat in the young system before decreasing slowly with $\Omega$, with the threat posed by plasma collection increasing with $\Omega$.

It is possible that a young fast rotating Sun could have had a rotation rate as high as $50~\Omega_\odot$.  It is  uncertain whether the Sun rotated that fast -- if that indeed occurred, it happened only for a short amount of time, relative to the Sun's lifetime, and at an age $\lesssim 100$~Myr. We simulated the Earth's magnetosphere in this extreme young system, modelling the wind of the Sun at speculative rotation rates of $30~\Omega_\odot$ and $50~\Omega_\odot$. For a fast-rotating Sun, our stellar wind models predict mass-loss rates of up to $10^{-11}~M_\odot$/yr, which agrees with observations of mass-loss rates derived for fast rotators, like AB Dor \citep{2019MNRAS.482.2853J}. However, our models do not consider wind saturation at very fast rotation. As a result, our wind models over-predict  angular momentum-loss at high rotation ($> 10~\Omega_\odot$). Given our choice of parameters, at rotation rates $30~\Omega_\odot$ and $50~\Omega_\odot$, our wind models have Mach numbers of 1.5 and 0.99, respectively. As a result, we expect Earth's surroundings to exhibit differences between the young system and the majority of its evolution. As the wind is sub-magnetosonic in the $50~\Omega_\odot$ model, in our speculative scenario, there would have been no bow shock surrounding the Earth in this fast young system. Once the rotation rate dropped to $30~\Omega_\odot$, a weak shock would then be formed, accompanied by a relatively inflated magnetosheath when compared to other models. With the Sun continuing to spin down, a strong shock would then surround our planet, and would remain for most of the duration of the solar main sequence.

\section*{Acknowledgements}
SC and AAV  acknowledge funding received from the Irish Research Council Laureate Awards 2017/2018. This work was carried out using the BATSRUS tools developed at The University of Michigan Center for Space Environment Modeling (CSEM) and made available through the NASA Community Coordinated Modeling Center (CCMC). The authors also wish to acknowledge the SFI/HEA Irish Centre for High-End Computing (ICHEC) for the provision of computational facilities and support. AAV thanks Dr.~Brian Wood for bringing to our attention the recent work on the moon regolith and its association to break in flare-CME relation for active stars, during the meeting ``The Solar and Stellar Wind Connection: Heating Processes and Angular Momentum Loss'' at the International Space Science Institute (ISSI). We thank the anonymous reviewer for their  constructive criticism, which helped improve the clarity of our paper.

\bibliographystyle{mnras}



\appendix

\section{Stellar wind fits}
 For ease of use in future works, here we provide the fit parameters obtained for some physical quantities of the solar wind at 1au, as a function of rotation rate. The fits are shown as solid/dashed lines in Figure \ref{Wind_4Panel} and take the form:
\begin{equation}\label{eq.fit}
    \log_{10}\big(F(\Omega)\big) = a\Bigg(\frac{\Omega}{\Omega_\odot}\Bigg)^b + c\Bigg(\frac{\Omega}{\Omega_\odot}\Bigg)^d + e\Bigg(\frac{\Omega}{\Omega_\odot}\Bigg)^f
\end{equation}
The functions $F(\Omega)$ are computed at 1 au, and are the following: The stellar wind radial and azimuthal velocities in km/s; stellar wind radial and azimuthal magnetic field strengths in G; stellar wind mass density in g/cm$^{-3}$, temperature in MK; mass-loss rate in M$_\odot$/yr and angular momentum loss rate in erg. The parameters $a$-$f$  for each of these functions are shown in Table \ref{table_fits1} for $\Omega < 1.4\Omega_\odot$ and \ref{table_fits2} for $\Omega \geq 1.4\Omega_\odot$.

\begin{table*}
\caption{The fitting parameters to derive the stellar wind properties at 1au for $\Omega < 1.4 \Omega_\odot$. Parameters $a$ to $f$ should be implemented in Equation (\ref{eq.fit}) to derive the solar wind conditions. }
\begin{tabular}{c c c c c c c c c}
\hline

	Parameter & $u_r^{\rm sw}$ & $u_\phi^{\rm sw}$ & $B_r^{\rm sw}$ & $B_\phi^{\rm sw}$ & $\rho^{\rm sw}$ & $T^{\rm sw}$ & $\dot{M}$ & $\dot{J}$ \\ \hline
	$a$ & -8.50$\times 10^{1}$ & -2.60$\times 10^{0}$ & -1.32$\times 10^{2}$ & 1.09$\times 10^{2}$ & -1.25$\times 10^{0}$ & -7.17$\times 10^{-2}$ & -1.21$\times 10^{0}$ & -5.71$\times 10^{2}$ \\
	$b$ & 5.84$\times 10^{-1}$ & -7.93$\times 10^{-1}$ & -2.32$\times 10^{-3}$ & 2.80$\times 10^{-1}$ & -2.93$\times 10^{0}$ & -2.80$\times 10^{0}$ & -2.95$\times 10^{0}$ & -1.85$\times 10^{-2}$ \\ 
	$c$ & 8.77$\times 10^{1}$ & 2.54$\times 10^{0}$ & 1.32$\times 10^{2}$ & -1.09$\times 10^{2}$ & 2.10$\times 10^{0}$ & 5.90$\times 10^{0}$ & -1.24$\times 10^{1}$ & 6.01$\times 10^{2}$ \\ 
	$d$ & 5.72$\times 10^{-1}$ & -1.82$\times 10^{0}$ & 2.02$\times 10^{-3}$ & 2.76$\times 10^{-1}$ & 3.37$\times 10^{-1}$ & 9.04$\times 10^{-2}$ & -1.03$\times 10^{-1}$ & -1.51$\times 10^{-2}$ \\ \hline
\end{tabular}
\label{table_fits1}
\end{table*}

\begin{table*}
\caption{The same as in Table \ref{table_fits1}, but for $\Omega \geq 1.4\Omega_\odot$.}
\begin{tabular}{c c c c c c c c c}
\hline
    Parameter & $u_r^{\rm sw}$ & $u_\phi^{\rm sw}$ & $B_r^{\rm sw}$ & $B_\phi^{\rm sw}$ & $\rho^{\rm sw}$ & $T^{\rm sw}$ & $\dot{M}$ & $\dot{J}$ \\ \hline
	$a$ & 8.21$\times 10^{-8}$ & 2.01$\times 10^{-1}$ & -1.32$\times 10^{2}$ & -9.56$\times 10^{1}$ & 1.83$\times 10^{0}$ & 1.22$\times 10^{2}$ & -1.26$\times 10^{1}$ & -5.71$\times 10^{2}$ \\ 
	$b$ & 3.91$\times 10^{0}$ & 6.71$\times 10^{-1}$ & -2.32$\times 10^{-3}$ & -6.97$\times 10^{-3}$ & 2.56$\times 10^{-1}$ & 9.22$\times 10^{-2}$ & -5.98$\times 10^{-2}$ & -1.85$\times 10^{-2}$ \\ 
	$c$ & 2.77$\times 10^{0}$ & -9.90$\times 10^{-1}$ & 1.32$\times 10^{2}$ & -1.33$\times 10^{-7}$ & -1.02$\times 10^{-2}$ & -1.16$\times 10^{2}$ & -1.24$\times 10^{-3}$ & 6.01$\times 10^{2}$ \\ 
	$d$ & 3.99$\times 10^{-2}$ & -3.32$\times 10^{-1}$ & 2.02$\times 10^{-3}$ & 3.79$\times 10^{0}$ & 1.42$\times 10^{0}$ & 9.52$\times 10^{-2}$ & 1.76$\times 10^{0}$ & -1.51$\times 10^{-2}$\\ 
	 $e$ & - & - & - & 9.59$\times 10^{1}$ & - & - & - & - \\ 
	 $f$ & - & - & - & 2.39$\times 10^{-3}$& - & - & - & - \\ 
	 \hline
\end{tabular}
\label{table_fits2}
\end{table*}

To find the solar wind conditions at other orbital distances $x$, we use the values obtained at Earth's orbit $x_E$ (i.e., from Tables \ref{table_fits1} and \ref{table_fits2}), with the following power-laws with distance
\begin{equation}
    B_r^{\rm sw}(x) = B_r^{\rm sw}(x_{E})  \bigg(\frac{x_{E}}{x}\bigg)^2,
\end{equation}
\begin{equation}
    B_\phi^{\rm sw}(x) = B_\phi^{\rm sw}(x_{E})  \bigg(\frac{x_{E}}{x}\bigg),
\end{equation}
\begin{equation}
    u_r^{\rm sw}(x) = u_r^{\rm sw}(x_{E}) ,
\end{equation}
\begin{equation}
    u_\phi^{\rm sw}(x) = u_\phi^{\rm sw}(x_{E})\bigg(\frac{x_{E}}{x}\bigg),
\end{equation}
\begin{equation}
    T^{\rm sw}(x) = T^{\rm sw}(x_{E}) \bigg(\frac{\rho(x)}{\rho(x_{E})}\bigg)^{\alpha-1} = T^{\rm sw}(x_E) \bigg(\frac{x_E}{x}\bigg)^{2(\alpha-1)},
\end{equation}
\begin{equation}
    \rho^{\rm sw}(x) = \rho^{\rm sw}(x_{E})\bigg(\frac{x_{E}}{x}\bigg)^2 ,
\end{equation}

\noindent where $\alpha = 1.05$ is the polytropic index. Note that in these relations, we assume that $u_r^{\rm sw}$ does not depend on $x$, which implies that the wind has reached terminal velocity. This is valid for orbital distances approximately larger than Mercury's orbit.

\section{Open and Closed Magnetospheres}\label{Zcomp}

\begin{figure*}
	\includegraphics[width=0.49\textwidth]{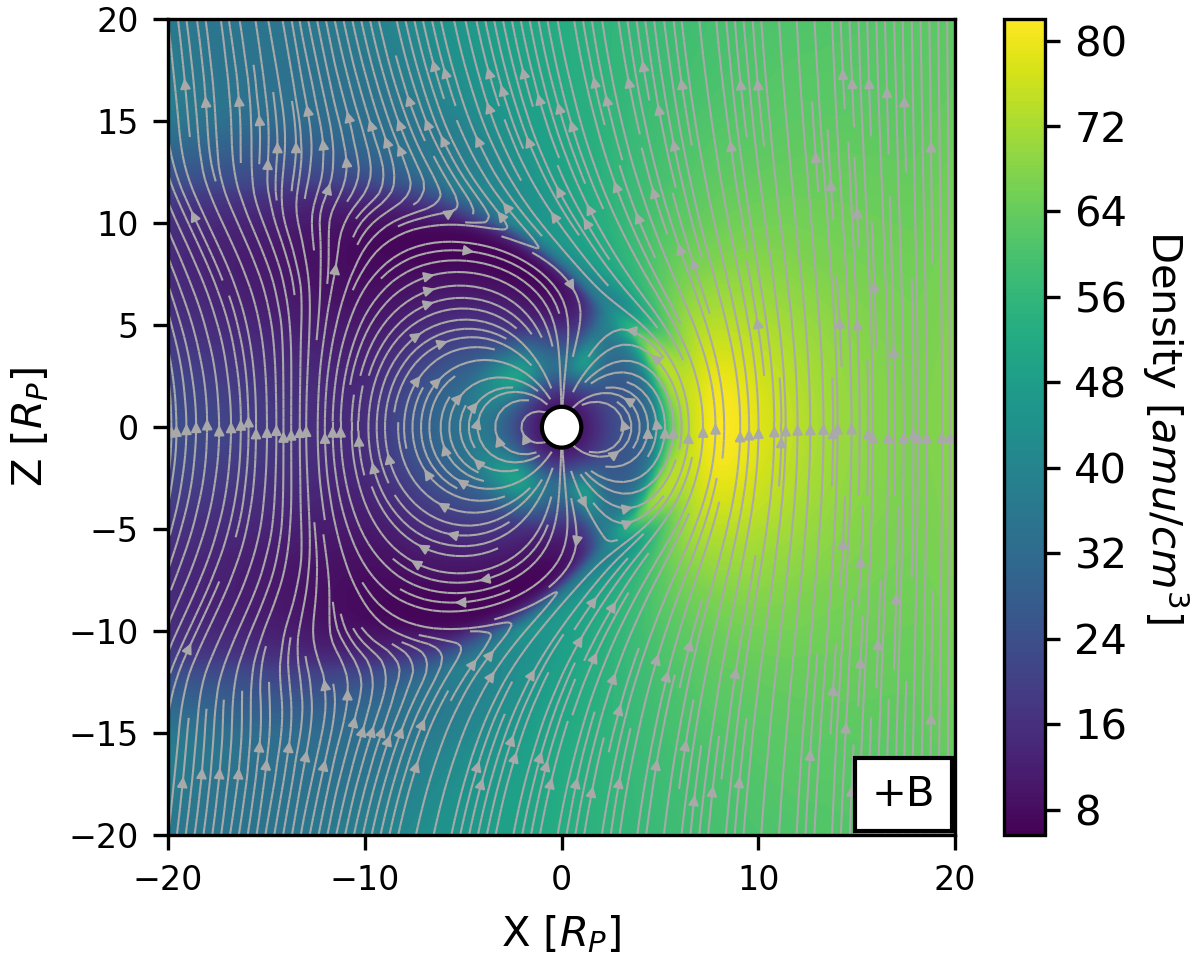}
	\includegraphics[width=0.49\textwidth]{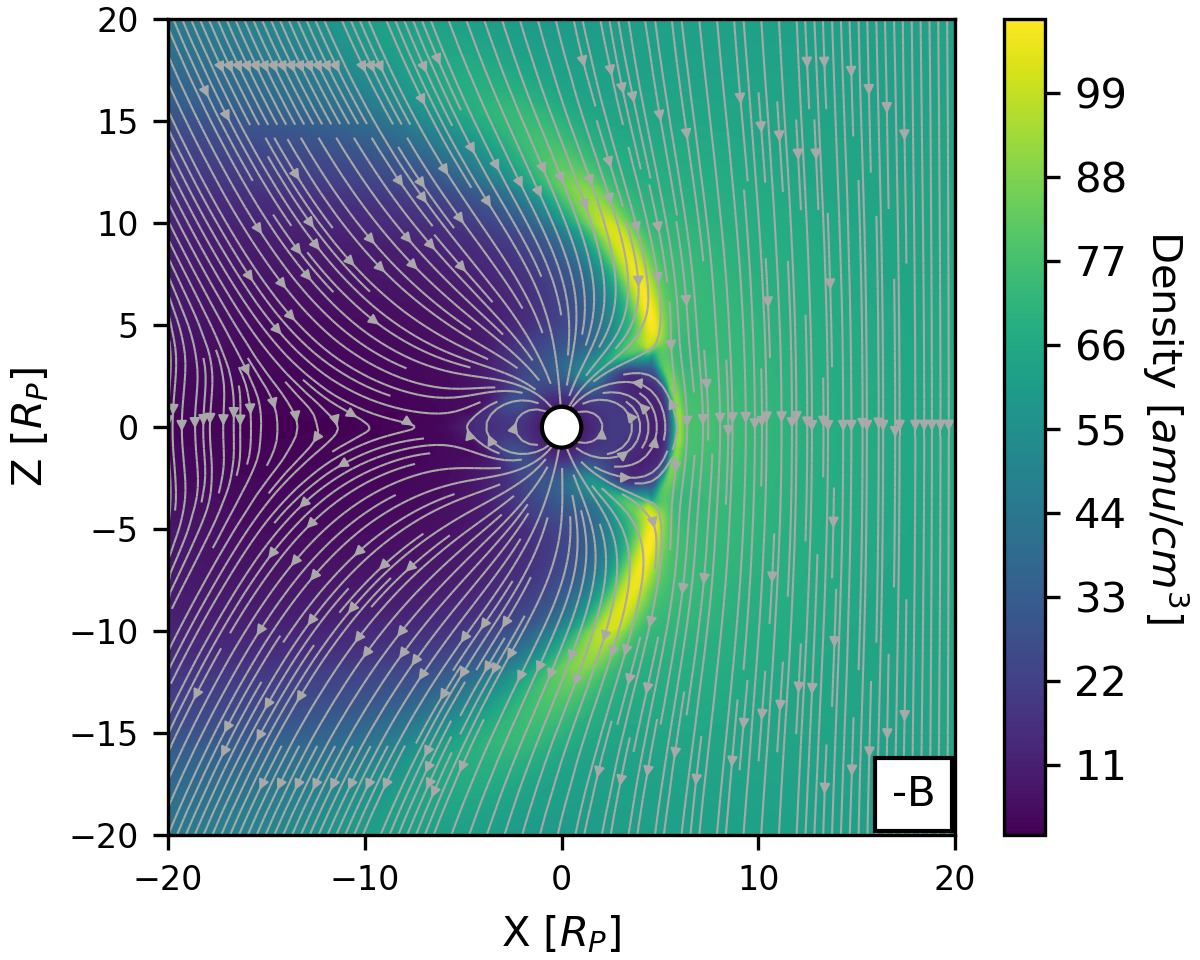}
	\includegraphics[width=0.49\textwidth]{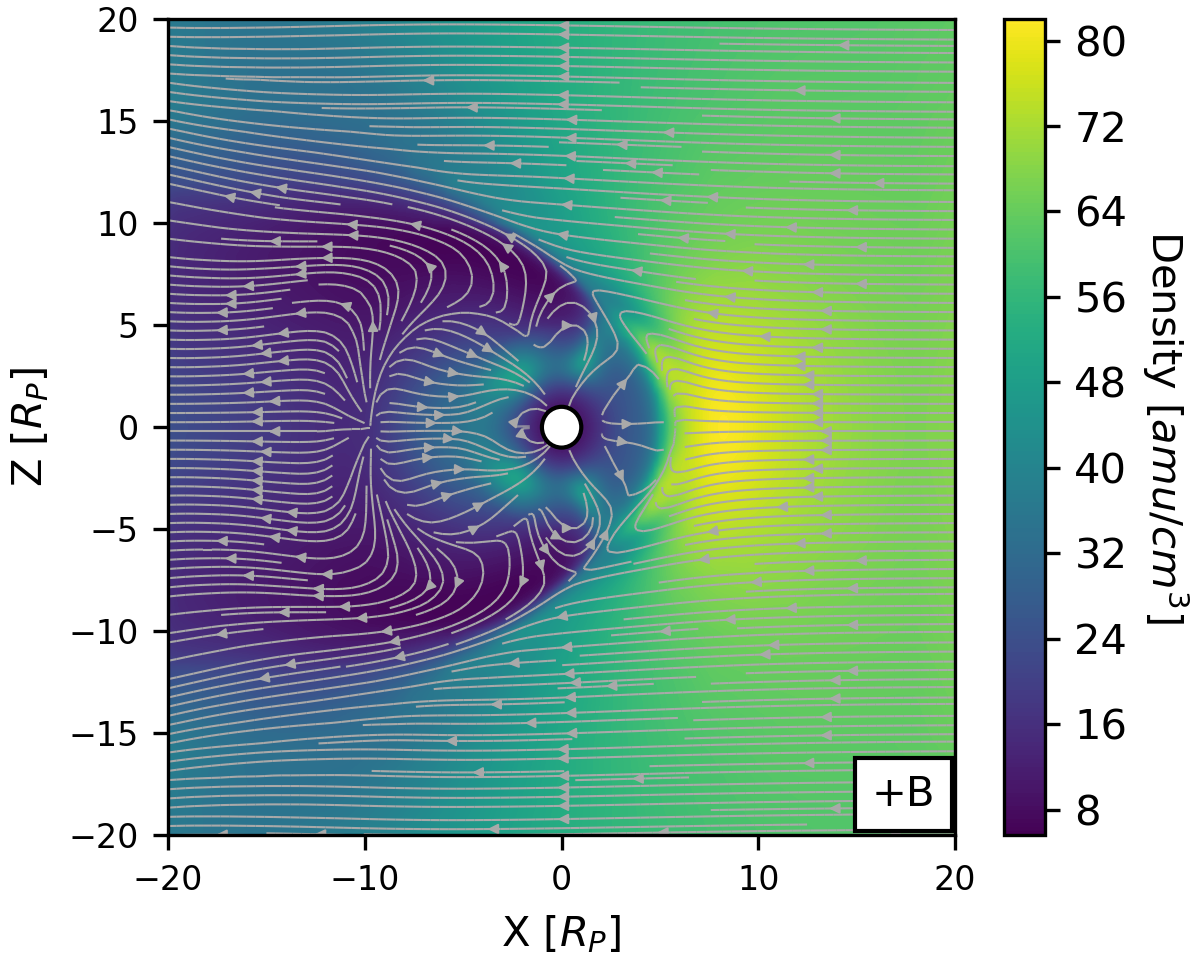}
	\includegraphics[width=0.49\textwidth]{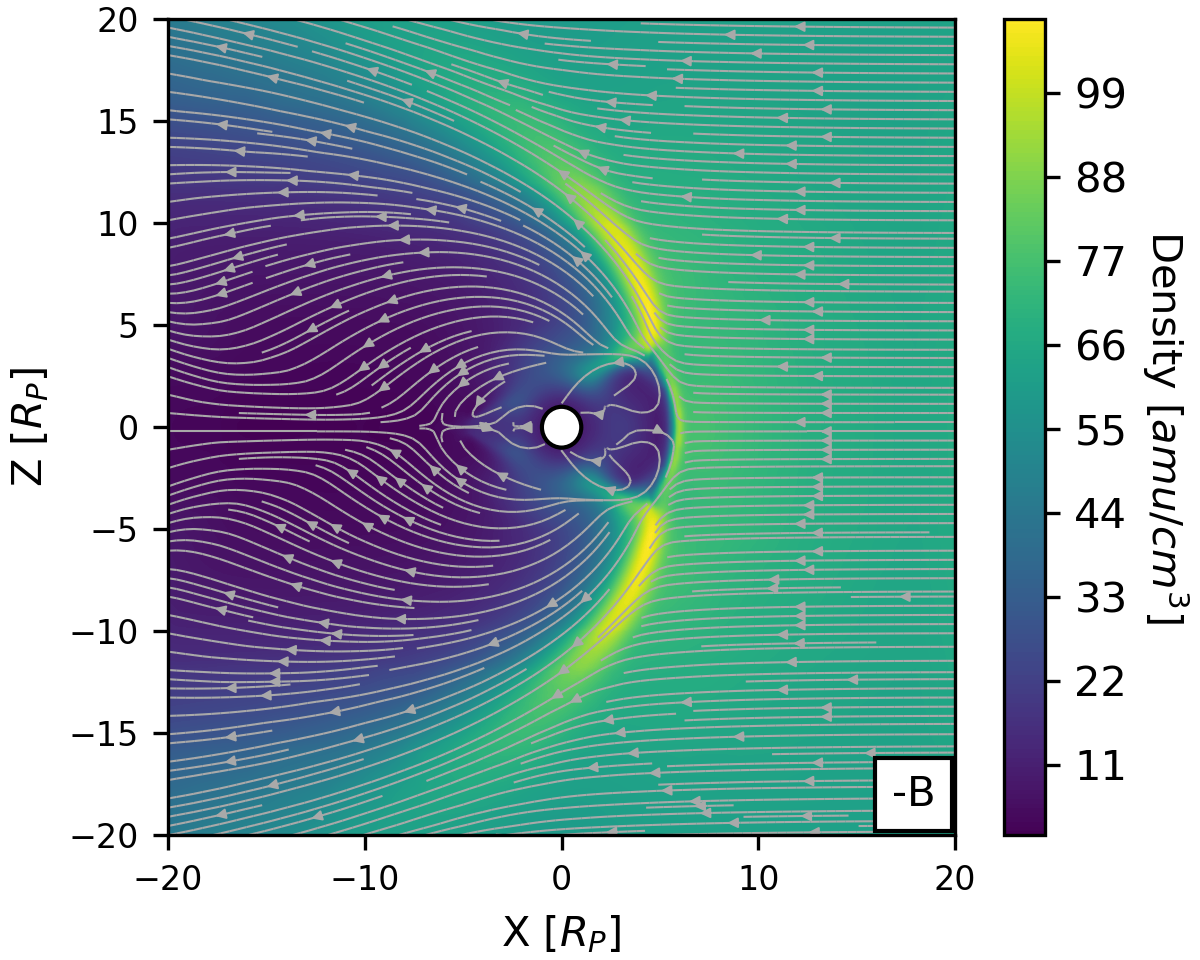}
    \caption{Closed (Left, northward stellar wind) and open (Right, southward stellar wind) magnetosphere. Top: The contour shows the density distribution in the models, while the streamtracers are magnetic field lines. Bottom: The contour shows the density distributions while the stream lines now show the velocity vectors. All parameters were kept constant between these models, except for the orientation of the stellar wind's magnetic field.}
    \label{fig_Zcomp_2d}
\end{figure*}

In Section \ref{section_EMIT}, we discussed the various trends observed in Earth's magnetosphere with time, from 1.5D stellar wind simulations. It is also likely that the polarity of either the star's or planet's magnetic field will flip cyclically. This may lead to both ``open" and ``closed" magnetospheres \citep{Das2018, Cravens2004Book}. To examine this difference we perform two simulations, both with the same parameters as the $1.2~\Omega_\odot$ wind (see Table \ref{wind-table}), but now with a northward and southward magnetic field (positive and negative Z). Instead of using a total magnetic field strength of $4.7\times10^{-5}~G$ as in Table \ref{wind-table}, we use a strength of $1\times10^{-3}~G$ so that the nightside magnetosphere is entirely contained in our numerical grid. These can be seen in Figure \ref{fig_Zcomp_2d}.

 For the positive case, the magnetic field is anti-aligned with the magnetic field of the planet. This creates a closed magnetosphere. For the negative case, the magnetic field in the wind is aligned with that of the planet. This creates an open magnetosphere. When the magnetic field lines of the wind and the magnetic field of the planet are anti-aligned, the planet's field lines remain closed on both the dayside and nightside of the planet, unlike what occurs in the models examined in Section \ref{section_EMIT}. This results in a reduced inflow of material from the stellar wind, due to the lack of open field lines around the poles. Magnetic reconnection no longer occurs on the nightside of the planet, which can drive material towards the atmosphere \citep{Das2018}. However due to an increase in closed field lines surrounding the planet, there is a larger amount of material held within these loops, compared to models in Section \ref{section_EMIT}.

When the fields are aligned an open magnetosphere is observed. There is now a much lower number of closed magnetic field lines, which are mostly on the dayside of the planet. Open field lines now exist at much lower latitudes than seen in other models -- this simulation gives a $\Phi = 23.3^{\circ}$, similar to the analytical value of $24.4^{\circ}$. As a result, there is now a much larger inflow of material, and a lower amount held by closed field lines. These two scenarios correspond to the two competing effects discussed by \citet{blackman2018}. In one case we have a much larger region covered by closed field lines and so a larger collecting area for plasma. While in the other we see a greater potential for inflow of material directly from the stellar wind.

\begin{figure*}
	\includegraphics[width=\columnwidth]{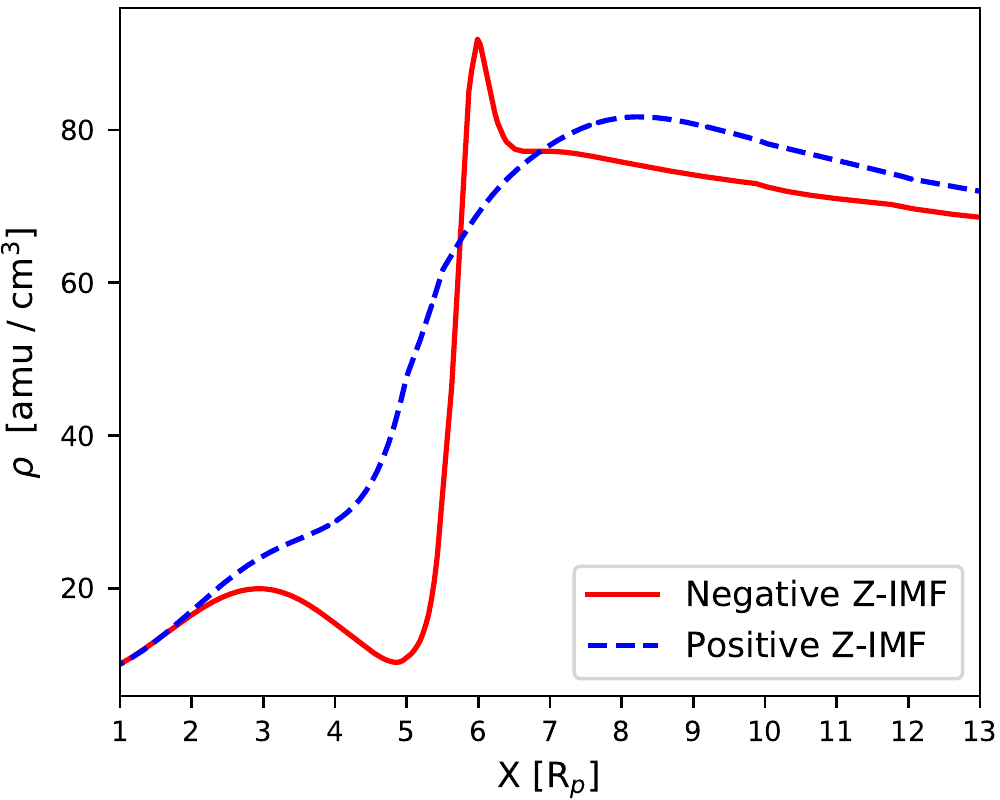}
	\includegraphics[width=\columnwidth]{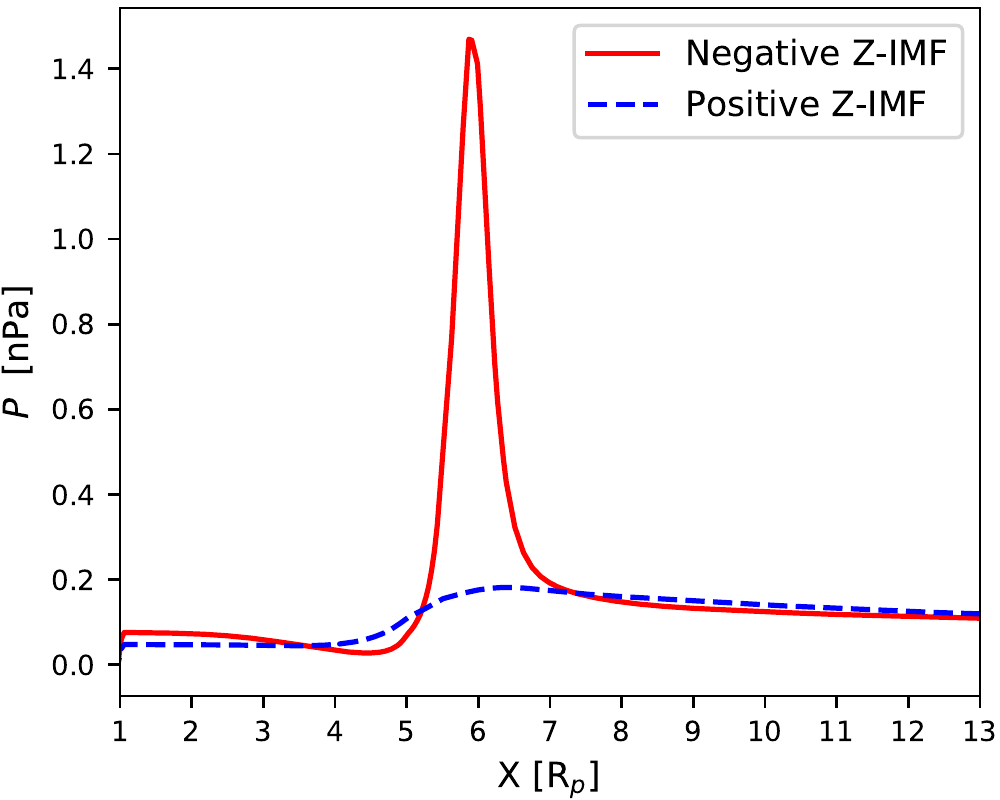}
    \caption{Left: The density distributions along the x axis for both a negative Z stellar wind (open) and a positive Z stellar wind (closed) case. Right: Variation in thermal pressure for both open and closed cases. We can see that the over density caused by magnetic reconnection on the dayside of the planet leads to a peak in thermal pressure.}
    \label{fig_Zcomp_rhoP}
\end{figure*}

As is clear in Figure \ref{fig_Zcomp_2d}, there are significant differences in the dayside of open and closed magnetospheres. To examine these differences, we look at the subsolar line, in a similar way to Section \ref{section_EMIT}. The density and thermal pressure distributions along this line can be seen in Figure \ref{fig_Zcomp_rhoP}.

In the open magnetosphere (Negative Z magnetic field) model, reconnection occurs on the dayside at a distance of $\sim5$~$R_P$ from the planet. This corresponds to both an under and over density seen in this model along the X-axis. On the planet side of this reconnection, material is driven away from this point with the closed field lines trapping a portion of this material. On the wind side of the reconnection, material is driven away from the planet but the stellar wind acts as a resisting force creating the overdensity seen in this model at $\sim6$~$R_P$. As a result there is an increase in thermal pressure at $\sim6$~$R_P$ as the forces from both reconnection and the stellar wind compress the material at this point.

In the closed magnetosphere there is a much smoother density profile on the dayside of the planet, as there is no reconnection on this side in the model. As a result the over and under densities observed in the open case are not seen here.  We no longer see a bow shock with high density that extends northward and southward, but instead the closed planetary magnetic field lines focus material towards the subsolar line. This can be seen in Figure \ref{fig_Zcomp_rhoP}, in the velocity streamlines. This forms a high density ``bubble" on the dayside instead of an extended shock. In the open model, this is not the case. The open planetary field lines from the planet lead to the opposite effect, where we observe more inflow of material, but also a more extended shock northwards and southwards.

\bsp	
\label{lastpage}
\end{document}